\documentclass[pdflatex,sn-mathphys-num]{sn-jnl}

\usepackage{graphicx}%
\usepackage{multirow}%
\usepackage{amsmath,amssymb,amsfonts}%
\usepackage{amsthm}%
\usepackage{mathrsfs}%
\usepackage[title]{appendix}%
\usepackage{xcolor}%
\usepackage{textcomp}%
\usepackage{manyfoot}%
\usepackage{booktabs}%
\usepackage{algorithm}%
\usepackage{algorithmicx}%
\usepackage{algpseudocode}%
\usepackage{listings}%
\usepackage{tabularx}
\usepackage{ragged2e}
\usepackage{wasysym}
\usepackage{float}
\usepackage{wrapfig}
\usepackage{makecell}
\usepackage[dvipsnames]{xcolor}
\usepackage{newtxtext}
\usepackage{newtxmath}
\usepackage{natbib}
\usepackage{hyperref}
\usepackage{gensymb}

\theoremstyle{thmstyleone}%
\theoremstyle{thmstyletwo}%

\theoremstyle{thmstylethree}%

\begin{document}

\title[Inertial forces and elastohydrodynamic interaction of spherical particles in wall-bounded sedimentation experiments at low particle Reynolds number]{Inertial forces and elastohydrodynamic interaction of spherical particles in wall-bounded sedimentation experiments at low $\textit{Re}_\text{P}$}


\author[1]{\fnm{Isabell} \sur{Noichl}}\email{isabell.noichl@mv.rptu.de}

\author*[1,2]{\fnm{Clarissa} \sur{Sch\"onecker}}\email{clarissa.schoenecker@h-da.de}

\affil*[1]{\orgname{RPTU University Kaiserslautern-Landau}, \orgaddress{\city{Kaiserslautern}, \postcode{D-67663}, \country{Germany}}}

\affil[2]{\orgname{Darmstadt University of Applied Sciences}, \orgaddress{\city{Darmstadt}, \postcode{D-64295}, \country{Germany}}}


\abstract{Unsteady, wall-bounded sedimentation of spheres at low particle Reynolds numbers, $Re_\text{P}\lesssim 0.1$, under the influence of small elastic deformation was investigated experimentally. The complete kinematics of elastic and rigid spheres sedimenting from rest at various initial distances from a rigid plane wall in a rectangular duct were measured. Several previously unrecognized phenomena arising from fluid inertia and superimposed elastohydrodynamic effects were identified and analyzed. Among these is an \textit{inertial wall attraction}, whereby particles migrate towards the wall during the initial acceleration phase. After this initial phase, rigid spheres sedimenting at $Re_\text{P}\approx O(10^{-1})$ followed behavior consistent with classic wall-lift models, including approximately linear migration away from the wall. In contrast, at smaller Reynolds numbers, $Re_\text{P}\approx O(10^{-2})$, both rigid and elastic spheres exhibited persistently unsteady sedimentation, characterized by deceleration despite increasing wall distance. These results enable the formulation of a conceptual framework that classifies near-wall sedimentation regimes according to particle Reynolds number and the position of boundaries relative to the Stokes length scale. For increasing deformability, the unsteady behavior was further modulated by nonlinearities. The observations suggest the presence of an \textit{elastohydrodynamic memory} effect arising from the coupling of fluid inertial forces with particle deformability. The experimental findings are supported by computational fluid dynamics simulations that provide qualitative insight into the evolving flow field. Overall, the results demonstrate that classic assumptions commonly applied to particle sedimentation in creeping flows break down in the presence of nearby boundaries and reveal a counterintuitive trend: as the particle Reynolds number decreases, fluid inertia can play an increasingly important role in governing particle motion near walls. The proposed conceptual framework may therefore aid the interpretation of the near-wall dynamics of deformable microplastic particles, for which comparable material properties and flow regimes are encountered in environmental and wastewater flows.}

\keywords{particle-laden flow, low Reynolds number flows, transient dynamics}



\maketitle

\section{Introduction} \label{sec1}
Particle-laden fluid flows are ubiquitous in our daily lives. From the flow of our blood cells through the cardiovascular system to the removal of microplastics from wastewater in sewage treatment plants---particles of every conceivable shape or morphology in interaction with fluids are found everywhere. Typically, these particle-laden fluids are within or flow through confined spaces, such as reactors, pipelines and tubes or blood vessels. In other words, the particles do not only interact with the (flowing) fluid but also with surrounding walls. These interactions of particles, the surrounding fluid and a confining wall can lead to phenomena, to which everyone of us is exposed to---perhaps without knowing it. A well-known example of such a phenomenon is the Fåhraeus effect. Robin Fåhraeus described in 1929 the effect, that blood cells migrate laterally from the walls of blood vessels in direction of the vessel axis. A cell-free plasma layer is formed along the vessel walls, which is of crucial importance for the functionality of the cardiovascular system and for immune defense.\cite{Fahraeus.1929} Today it is known, that the Fåhraeus effect in blood vessels is subject to the principle of elasto-inertial focusing, which is a combined effect of inertial focusing and elastic forces.\newline 
The part of elastic forces can originate either from elasticity of the particle, the walls or from viscoelastic properties of the fluid.\cite{Naderi.2022} Segré and Silberberg were the first researchers, who, in 1961, reported observations on the phenomenon of particle collection of rigid particles in viscous fluids due to fluid inertia. In their experiments, they observed radial particle displacements in Poiseuille flows of suspensions. Today known as Segré-Silberberg effect, it describes that macroscopic, rigid and spherical particles collect into a thin, annular region when an initially uniform dilute suspension is passed in laminar flow through a straight cylindrical tube.\cite{Segre.1961} Segré and Silberberg recognized the potential of this effect as particle sorting mechanism. In recent years, the principle of inertial focusing has attracted increasing interest in the field of microfluidics.\cite{Naderi.2022}, \cite{DiCarlo.2007}, \cite{Zhou.2020}, \cite{Choi.2020} This interest is based mainly on the numerous possible applications in, for example, the field of biomedical diagnostics. For example, it is well-known, that infectious diseases like COVID-19 can alter the physical properties of blood cells, including morphological or mechanical features.\cite{Kubankova.2021} For this reason, deformability cytometry techniques are promising in the field of real-time diagnostics but also in the field of material characterization of deformable micro-particles.\cite{Mietke.2015} What the focusing techniques from the field of inertial microfluidics have in common: the word “inertial” refers mostly to separately imposed advective fluid inertia, \textit{i.e.}, the particles are moved by and with the fluid.\newline
\indent The lateral migration of particles perpendicular to walls like in the Segré-Silberberg effect is due to a combination of advective inertia induced lift forces directed towards the wall and wall-induced lift forces directed away from the wall. The former are induced by shear gradients, \textit{i.e.}, the particle experiences locally a shearing due to gradients in the velocity (\textit{cf.} parabolic flow profiles in cylindrical tubes). The mechanism was first described by P.G. Saffman in 1965.\cite{Saffman.1965} The latter lift forces are induced solely by the presence of the wall and can also emerge if the fluid is quiescent (not separately advected).\newline
\indent The time reversal symmetry of the steady Stokes equations and the symmetry in front of and behind the point at the surface closest to the wall would actually imply the absence of normal forces in the creeping flow regime.\cite{Bureau.2023} This would be especially true if both the fluid Reynolds number and the particle Reynolds number are close to zero ($Re_\text{P}\ll1$), \textit{i.e.,} if the particle itself translates with small mass inertia and momentum transferred to the fluid is very small. But why do normal forces like wall-induced hydrodynamic lift emerge anyway? The simple answer is: inertia of the surrounding fluid is per se finite. In that moment, in which a particle starts translating through a fluid, a disturbance flow is generated and the fluid around the particle accelerates. In this case, Stokes's assumptions of reversibility are no longer valid in the far-field of the particle and the equations of motion must be supplemented by the influence of fluid inertia as recognized by Carl Wilhelm Oseen in 1910.\cite{Oseen.1910} The inertial contribution is indeed small for low Reynolds numbers. However, if a sphere translates in the presence of a wall, these small inertial contributions are enough to induce a lift force directed normally to the wall due to symmetry-breakings in the disturbance flow. It was Carl Wilhelm Oseen´s PhD student, Hilding Faxén, and later Vasseur and Cox who described the lateral migration of a rigid particle in a viscous fluid and in the presence of plane walls theoretically.\cite{Faxen.1922b}, \cite{Vasseur.1977b} These theoretical approaches refer purely to rigid and perfectly spherical particles with small but finite inertia. However, there are several other effective mechanisms than pure fluid inertia that can invoke the symmetry breaking that leads to lift forces. Such mechanisms are, \textit{e.g.,} elastic deformation of the particle or of the confining wall. These deformations also induce a symmetry-breaking, which leads to so called elastohydrodynamic lift forces.\cite{Urzay.2007}, \cite{Bertin.2022} Theoretical work on lift forces in the presence of a wall usually assume a steady motion of the object, or steady relative velocity, respectively.\cite{Bertin.2022}, \cite{Ekanayake.2021}. Furthermore, experimental and theoretical work on particle dynamics including elastohydrodynamic effects is mostly performed with small gap sizes between the particle and the wall, or the elastic layer, respectively, to make use of, \textit{e.g.} elastohydrodynamic lubrication theory.\cite{Beaucourt.2004}, \cite{Davies.2018}, \cite{Rallabandi.2017}, \cite{Rallabandi.2018}, \cite{Zhang.2020}\newline
\indent If the active motion of small particles in a viscous fluid is assumed to be governed by the steady Stokes equations, the hydrodynamics after self-propulsion can be described theoretically using multipole expansions of fundamental singularities. A specific class of microswimmers studied in this context are so called squirmers, which propel themselves by generating tangential surface deformations and oscillations by cilia motion.\cite{Lighthill.1952}, \cite{Blake.1971b} The leading-order singularity is the Stokeslet, which represents the fundamental solution of Stokes equations of the flow induced by a point force in the fluid. Higher-order singularities, \textit{i.e.} derivatives of the Stokeslet, include force dipoles (the Stresslet) and the source dipole. The presence of boundaries, such as walls or interfaces, alters these singularity solutions due to no-slip or stress conditions.\newline
\indent The principle of hydrodynamic imaging refers to the remote sensing of objects based on disturbances generated by moving or inert particles in a flow field.\cite{Takagi.2020} This principle is applied in modeling flow fields near walls by introducing fictitious singularities that enforce the required boundary conditions, analogous to the method of electrical images in electrostatics.\cite{Blake.1971}, \cite{Blake.1974}, \cite{Spagnolie.2012} In fact, theoretical squirmer dynamics are strongly influenced by walls, \textit{e.g.}, walls modify the propulsion efficiency, the squirmer´s orientation and the stability near surfaces.\cite{Lintuvuori.2016}, \cite{Ruhle.2018}, \cite{Tingting.2023}, \cite{Thery.2023} Many experimentally observed phenomena can be constructed and simulated using the squirmer model and multipole analysis near walls. Nevertheless, it is assumed that the particles translate with steady velocity, the flow field generated by the singularities is instantaneous and---except for gravity---most models do not account for body forces.\newline
\indent In reality, however, particles with finite mass and finite body size also undergo transient phases of motion, \textit{e.g.}, through spontaneous, time-depended self-propulsion or during the transient phase of gravitational acceleration like in sedimentation processes.\cite{Lovalenti.1993}, \cite{Redaelli.2022}, \cite{Noichl.2022} During transient phases of particle motion, unsteady forces related to inertia of mass act on a particle. An example for such an unsteady force acting on a particle during mass acceleration from rest is the Basset history force classifying as a so called memory effect.\cite{Basset.1888}, \cite{Feng.1995} This type of inertial force is particularly relevant in an otherwise quiescent liquid when mass inertia of the particle is small but finite. As a consequence, steady-state conditions are no longer valid and the complete transient dynamics or kinematics must be taken into account.\newline
\indent Experimental investigations on the complete kinematics are rare and, especially, investigations of single elastic spheres sedimenting in the Reynolds number regime $Re_\text{P}\lessapprox 0.1$ and in which the influence of bounding plane walls is precisely addressed are not available to date. Complete kinematics in this context includes the spatio-temporal resolution of the entire trajectory from the beginning of the acceleration, \textit{e.g.,} when starting from rest. Fully resolved kinematics were reported, for example, for rigid particles sedimenting at moderate particle Reynolds numbers (however, $Re_\text{P}\gtrapprox0.1$), for rising bubbles in viscous liquids or for sedimentation of rigid spheres in a cylindrical tube filled with a viscoelastic liquid.\cite{Jaroslawski.2025}, \cite{Takemura.2002}, \cite{Becker.1994} Generally, in the past the light was often directed towards the description of drag and lift forces of drops and bubbles.\cite{Magnaudet.2003} However, the underlying physics of bubbles and drops in liquids cannot be easily applied to the interaction of deformable solids with liquids. All the gaps and limitations with respect to transient dynamics in the literature prevent detailed insights into particle dynamics which are highly relevant for numerous applications, \textit{e.g.,} the treatment of microplastics-contaminated wastewater or particle sedimentation along surfaces coated by biofilms.\cite{Aurich.2023} One possible reason for the experimental gaps could be that it is difficult to spatially and temporarily resolve the kinematic phenomena in the µm-range and especially in the near-wall region, where very small Reynolds numbers $Re_\text{P}\lessapprox 0.1$ naturally occur. \newline
\indent In this report, we therefore attempt to shed light on the phenomena of transient particle dynamics with and without elastohydrodynamic wall interactions using an experimental approach. The focus is on spontaneous particle sedimentation from rest in an initially quiescent fluid within a rectangular duct. Both elastic and rigid spheres were examined, which started at various distances from rigid walls. The experiments were performed in a scaled experimental design in the cm-range. The advantage of this scaling is that the kinematics could be detected much more precisely in space and time than this would be possible on the µm-scale. The paper is structured as follows: section \ref{sec2} reviews the fundamental theory of classic Newtonian particle dynamics in fluids to calculate relevant forces and velocities of a flowing, single particle at low Reynolds numbers. Furthermore, the applicability and the limitations of existing theory are discussed. In section \ref{sec3}, the experimental methods such as the fabrication of polymer spheres and the optical measurement methods for tracking the particle kinematics are explained. In section \ref{sec4}, data of experiments with rigid and elastic model particles sedimenting at different particle Reynolds numbers at various distances to rigid walls are presented. Additionally, 3D computational fluid dynamics (CFD) studies of a falling sphere in a rectangular fluid domain were carried out to gain insights into the principal fluid flow field during the experiment. Section \ref{sec5} summarizes the results from the experiments and simulations. Finally, the various types of inertial forces involved in the dynamical system are reviewed and possible mechanisms leading to the observed kinematics are discussed.\newline
\section{Theoretical background} \label{sec2}
In the following, a brief overview of the basic equations, which form the basis for the physical understanding of the data shown in the next sections, is given.\newline
\indent The unsteady equation of motion of a rigid, spherical particle in a spatially nonuniform, time-dependent fluid flow in the absence of any rigid boundaries, \textit{e.g.}, walls or other particles, is described by the Maxey-Riley (MR) equation.\cite{Maxey.1983}
\begin{equation} 
\label{eq1}
\begin{split}
 m_\text{P} \frac{\text{d} \textbf{\textit{U}}_\text{P}}{\text{d}t}= \left(m_\text{P}-m_\text{f}\right) & \textbf{\textit{g}}+ m_\text{f} {\left.\frac{\text{D}\textbf{\textit{U}}_\text{f}}{\text{D}t} \right|}_{{\textbf{\textit{X}}_\text{P} \left(t\right)}}-\frac{1}{2} m_\text{f} \frac{\text{d}}{\text{d}t} {\left.\left(\textbf{\textit{U}}_\text{P}-\textbf{\textit{U}}_\text{f}-\frac{R^2}{10}\nabla^2 \textbf{\textit{U}}_\text{f} \right) \right|}_{{\textbf{\textit{X}}_\text{P} \left(t\right)}}\\
& -6 \pi \eta R{\left.\left(\textbf{\textit{U}}_\text{P}-\textbf{\textit{U}}_\text{f}-\frac{R^2}{6}\nabla^2 \textbf{\textit{U}}_\text{f} \right) \right|}_{{\textbf{\textit{X}}_\text{P} \left(t\right)}}\\
& -6 \pi \eta R^2 \int_{0}^{t} \frac{\text{d}\tau}{\sqrt{\pi \nu \left(t-\tau \right)}} \frac{\text{d}}{\text{d}\tau} {\left.\left(\textbf{\textit{U}}_\text{P}-\textbf{\textit{U}}_\text{f}-\frac{R^2}{6}\nabla^2 \textbf{\textit{U}}_\text{f} \right) \right|}_{{\textbf{\textit{X}}_\text{P} \left(\tau\right)}}
\end{split}
\end{equation}
The Maxey-Riley equation results from the force balance around a sphere with mass $m_\text{P}=\frac{4}{3}\pi R^3\rho_\text{P}$. $R$ is the radius of the spherical particle and $\rho_\text{P}$ is the solid density of the particle. The sphere starts from rest and translates through an incompressible, Newtonian fluid with density $\rho$ and dynamic viscosity $\eta$. $\nu=\eta/\rho$ is the kinematic viscosity of the fluid. $m_\text{f}=\frac{4}{3}\pi R^3\rho$ is the mass of the fluid displaced by the sphere. The particle translates with velocity $\textbf{\textit{U}}_\text{P} =\frac{\text{d} {{\textbf{\textit{X}}_\text{P} \left(t\right)}}}{\text{d}t}=\begin{pmatrix} U_\text{P} & U_\text{P,y} & U_\text{P,z}\end{pmatrix}^T$ in a Lagrangian reference frame through the fluid in which $\textbf{\textit{X}}_\text{P} \left(t\right)$ is the position of the particle center at a certain time $t$. The fluid motion satisfies the incompressible Navier-Stokes equations whose solution is an Eulerian velocity field $\textbf{\textit{U}}_\text{f}=\textbf{\textit{u}}_\text{f}\left(\textbf{\textit{X}}_\text{P} \left(t\right),t\right)$ from the point of view of the particle center. This velocity field is the net velocity field of, on the one hand, a global background flow which is imposed independently of the particle's motion. An example for such an imposed flow is a Poiseuille flow through a tube. On the other hand, the velocity $\textbf{\textit{U}}_\text{f}$ at time $t$ contains the local disturbance flow around the sphere caused by the previous motion of the sphere. The term on the left side of Eq. \ref{eq1} is the inertia of the spherical particle due to acceleration of the particle´s mass $\textbf{\textit{F}}_\text{I}^P$. The mass inertia is balanced by the gravitational force $\textbf{\textit{F}}_\text{G}$, buoyancy $\textbf{\textit{F}}_\text{Buoyancy}$, advective fluid inertia $\textbf{\textit{F}}_\text{I}^f$, added mass $\textbf{\textit{F}}_\text{AM}$, viscous drag $\textbf{\textit{F}}_\text{D}^{ub}$ and the Basset history force $\textbf{\textit{F}}_\text{B}$, respectively (terms on the right side of Eq. \ref{eq1} in the order of appearance). If the fluid is quiescent, unbounded, and $Re\rightarrow 0$ ($\textbf{\textit{U}}_\text{f}= \begin{pmatrix} 0 & 0 & 0\end{pmatrix}^T$), Eq. \ref{eq1} reduces to the Basset-Boussinesq-Oseen (BBO) equation, in which all terms in Eq. \ref{eq1} containing $\textbf{\textit{U}}_\text{f}$ disappear.\cite{Basset.1888}, \cite{Boussinesq.1903}, \cite{Oseen.1927} The terms of order $R^2\nabla^2 \textbf{\textit{U}}_\text{f}$ in Eq. \ref{eq1} which arise in the added mass term, the viscous drag and the Basset history force are known as Faxén terms since they were originally derived by Hilding Faxén in his dissertation.\cite{Happel.1983} The so called Faxén´s law reads 
\begin{equation}
\label{eq2}
\textbf{\textit{F}}_\text{D}^{ub}=-6\pi \eta R{\left.\left(\textbf{\textit{U}}_\text{P}-\textbf{\textit{U}}_\text{f}-\frac{R^2}{6}\nabla^2 \textbf{\textit{U}}_\text{f} \right) \right|}_{{\textbf{\textit{X}}_\text{P} \left(t\right)}}
\end{equation}
and modifies Stokes´s drag force, or Stokes´s law for viscous drag on a sphere, respectively, with an additional fluid inertia term that takes into account the curvature (\textit{i.e.} the acceleration) of the disturbance flow induced by the sphere.\cite{Stokes.1851}, \cite{Maxey.1983} If there is a background flow involved in addition to the disturbance flow, the fluid inertia term $m_\text{f} {\left.\frac{\text{D}\textbf{\textit{U}}_\text{f}}{\text{D}t} \right|}_{{\textbf{\textit{X}}_\text{P} \left(t\right)}}$ plays a significant role in the Maxey-Riley equation. $\frac{\text{D}}{\text{D}t}\equiv\frac{\partial}{\partial t}+\left(\textbf{\textit{U}}_\text{f}  \cdot \nabla \right)$ is the material derivative in the background flow. The term accounts for advective inertia due to fluid motion acting on the sphere. The Maxey-Riley equation in this form is only valid, when the particle size is small compared to the characteristic scales of the spatial variations of the undisturbed background flow. For the case when the size of the particle is appreciable relative to characteristic length scale of the flow, Rallabandi extended the fluid inertia term to include the influence of curvature in the nonuniform background flow on advective inertia.\cite{Rallabandi.2021} Such scenarios can be found, for example, in microfluidic channels, where vortex formation due to cavities, takes place and where particles move within these vortices.\cite{Haddadi.2017}\newline
\indent As mentioned previously, both the BBO equation and the MR equation are valid only when a rigid, spherical particle is moving in an unbounded fluid. The influence of walls on the steady dynamics of spherical particles suspended in initially quiescent fluids or in (non)uniform flows is an ongoing part of intensive research for many years. When a particle translates in the vicinity of a wall and when both mass inertia and fluid inertia are present, symmetry breaking due to the wall leads to an increased drag $\textbf{\textit{F}}_\text{D}^{wb}$ and a wall-induced lift force $\textbf{\textit{F}}_\text{L}^{wi}$. However, analytical expressions for the drag force in wall-bounded fluids $\textbf{\textit{F}}_\text{D}^{wb}$ or the wall-induced hydrodynamic lift force $\textbf{\textit{F}}_\text{L}^{wi}$ are described only for several special cases, \textit{e.g.}, for one plane wall, linear shear flows, etc. A good overview of wall corrections for stationary drag and for wall lift models is given by Shi and Rzehak (2020) or by Ekanayake \textit{et al.} (2021).\cite{Shi.2020}, \cite{Ekanayake.2021}\newline
\indent The wall-induced hydrodynamic lift force as a function of the lift coefficient $c_\text{L}$ can be expressed by 
\begin{equation}
    \label{eq3}
F_\text{L}^{wi}=c_\text{L}R^2 \rho U_\text{P}^2 .
\end{equation}
Ekanayake \textit{et al.} (2021) developed a generalized lift model valid for arbitrary distances $d$ between the particle and the wall and proposed the following lift coefficient $c_\text{L}$ for non-rotating particles moving with $Re_\text{P}\leq 0.05$ in a quiescent fluid parallel to a plane wall:\cite{Ekanayake.2021}
\begin{equation}
    \label{eq4}
c_\text{L}=\frac{18\pi}{32+2\left(\frac{d}{L_\text{S}}\right)+3.8\left(\frac{d}{L_\text{S}}\right)^2+0.049\left(\frac{d}{L_\text{S}}\right)^3}+0.4353\left(\frac{R}{d}\right)-1.198\left(\frac{R}{d}\right)^2+0.7792\left(\frac{R}{d}\right)^3.
\end{equation}
The first fractional term corresponds to the lift coefficient in Takemura's outer-region model, where $L_\text{S}=\eta/\rho U_\text{P}$ is the Stokes length scale.\cite{Takemura.2004} \newline
\indent What existing models for wall-induced lift forces on spherical particles in quiescent liquids without imposed fluid flow (absence of shear) have in common: the wall-induced lift force is always positive and directed perpendicularly to the wall, \textit{i.e.}, it leads to a migration of the sphere away from the wall.\newline
\indent In 1922, Faxén was the first who derived an equation for the steady drag force when a spherical particle is sedimenting in the presence of a plane wall. He used his previously formulated unbounded drag force $\textbf{\textit{F}}_\text{D}^{ub}$ from Faxén´s law (Eq. \ref{eq2}) to calculate the influence on drag due to the disturbance flow induced by the sphere which is then reflected at the plane wall. This reflection in turn causes a background flow which affects the resulting stationary force on the sphere leading to an expression for the drag force in a bounded fluid $\textbf{\textit{F}}_\text{D}^{wb}$. Using Faxén´s $\textbf{\textit{F}}_\text{D}^{wb}$, the corrected Stokes´s velocity near a single plane wall, \textit{i.e.}, the dimensionless velocity $U_{\text{P}}/U_{\text{St}}$ in direction of gravitational acceleration, can be calculated as follows.\cite{Faxen.1922b}, \cite{Happel.1983}
\begin{equation}
    \label{eq5}
{\left(\frac{U_\text{P}}{U_\text{St}}\right)}_{\text{F}, \left.\bullet \right|}=1-\frac{9}{16}\left(\frac{R}{d}\right)+\frac{1}{8}\left(\frac{R}{d}\right)^3-\frac{45}{256}\left(\frac{R}{d}\right)^4-\frac{1}{16}\left(\frac{R}{d}\right)^5
\end{equation}
The index “$\left. \bullet \right|$” illustrates that it is the corrected dimensionless velocity for a sphere in the close vicinity of one plane wall. In the other spatial directions, the fluid is unbounded. The velocity $U_\text{St}$ results from Stokes´s law and is the theoretical velocity of a sphere which falls stationary due to gravity in an unbounded, incompressible fluid.\cite{Stokes.1851} $U_\text{St}$ results from balancing Stokes´s drag, gravity and buoyancy and reads
\begin{equation} \label{eq6}
U_\text{St}=\frac{2R^2 \Delta \rho g}{9\eta} .
\end{equation}
Thus, with the dimensionless velocity $U_{\text{P}}/U_{\text{St}}$ from Eq. \ref{eq5}, the influence of the wall on the steady velocity can be read directly. Faxén´s Eq. \ref{eq5} is only valid in the creeping flow regime when $Re_\text{P}\rightarrow0$. If the Reynolds number is slightly larger, \textit{e.g.}, of $O(10^{-1})$, the theoretical (dimensionless) velocities must be extended by the so-called Oseen correction terms, \textit{cf.} Eq. \ref{eq7} in section \ref{sec4_1} and Eq. \ref{eqA1} in Appendix~\ref{secA1_1} (the equations are explained in connection with the respective results for a rigid sphere sedimenting at $Re_\text{P}\approx O(10^{-1})$). \newline
\indent The discussed corrections are valid within the previously mentioned restrictions which are, on the one hand, rigidity and sphericity. On the other hand, the correction factors were derived by the assumption of stationary motion in the creeping flow regime. These assumptions help in theory to make particle dynamics accessible by various mathematical methods, \textit{e.g.}, series expansions. However, these assumptions represent reality in only very limited and isolated cases. There is especially little knowledge and experimental work available which deals with the influence of the presence of walls during the unsteady motion of particles, \textit{e.g.}, when they accelerate from rest due to their inertia of mass.\cite{Happel.1983}, \cite{Wakiya.1964} Furthermore, common theories about the influence of elasticity of the particles or the walls, respectively, on the dynamics are available only for certain exceptional cases. One example for such a case is the steady motion of an elastic particle in an unbounded fluid.\cite{Murata.1980b}, \cite{Villone.2019} For example, Murata expanded the theoretical Stokes´ drag around a perturbation parameter $\alpha=U_\text{St} \eta G^{-1} R^{-1}$ to derive the influence of small elastic deformation on the sedimentation velocity. $\alpha$ relates the viscous stresses to the elastic, mechanical shear stresses. If $\alpha$ is much smaller than unity, small deformations ($\Delta R\approx \alpha R$) are assumed. $G$ denotes the shear modulus of the elastic material with $G= E[2(1+\mu)]^{-1}$. $E$ represents Young´s elastic modulus and $\mu$ the Poisson´s ratio of the elastic solid. However, it was shown that Murata´s theory cannot be applied on sedimentation in bounded fluids, \textit{i.e.} measured sedimentation velocities of soft spheres in the center of a rectangular duct were significantly larger than predicted by Murata´s theory. Nevertheless the theory correctly predicts an increase in velocity since it predicted the deformation into a prolate spheroid for the corresponding solid and liquid material properties.\cite{Noichl.2022} Another case, or configuration, respectively, for which theoretical models with limitations exists, is the already mentioned translation of an object at very small distances between the elastic object and/or the elastic wall or a membrane. In this case, elastohydrodynamic lubrication theory is applicable, \textit{i.e.}, the Reynolds equations for thin fluid films hold.\cite{Weekley.2006}, \cite{Urzay.2007}, \cite{Rallabandi.2017}, \cite{Rallabandi.2018}, \cite{DaddiMoussaIder.2018}, \cite{Bertin.2022} In real systems, however, a complex interplay of the various effects like wall interaction at larger distances, elastohydrodynamic interaction and inertial forces takes place and cannot be considered in isolation from one another.\newline
\section{Methods} \label{sec3}
\subsection{Experimental setup and particle tracking velocimetry} \label{sec3_1}
\begin{figure}
    \centering
    \includegraphics[width=0.9\textwidth]{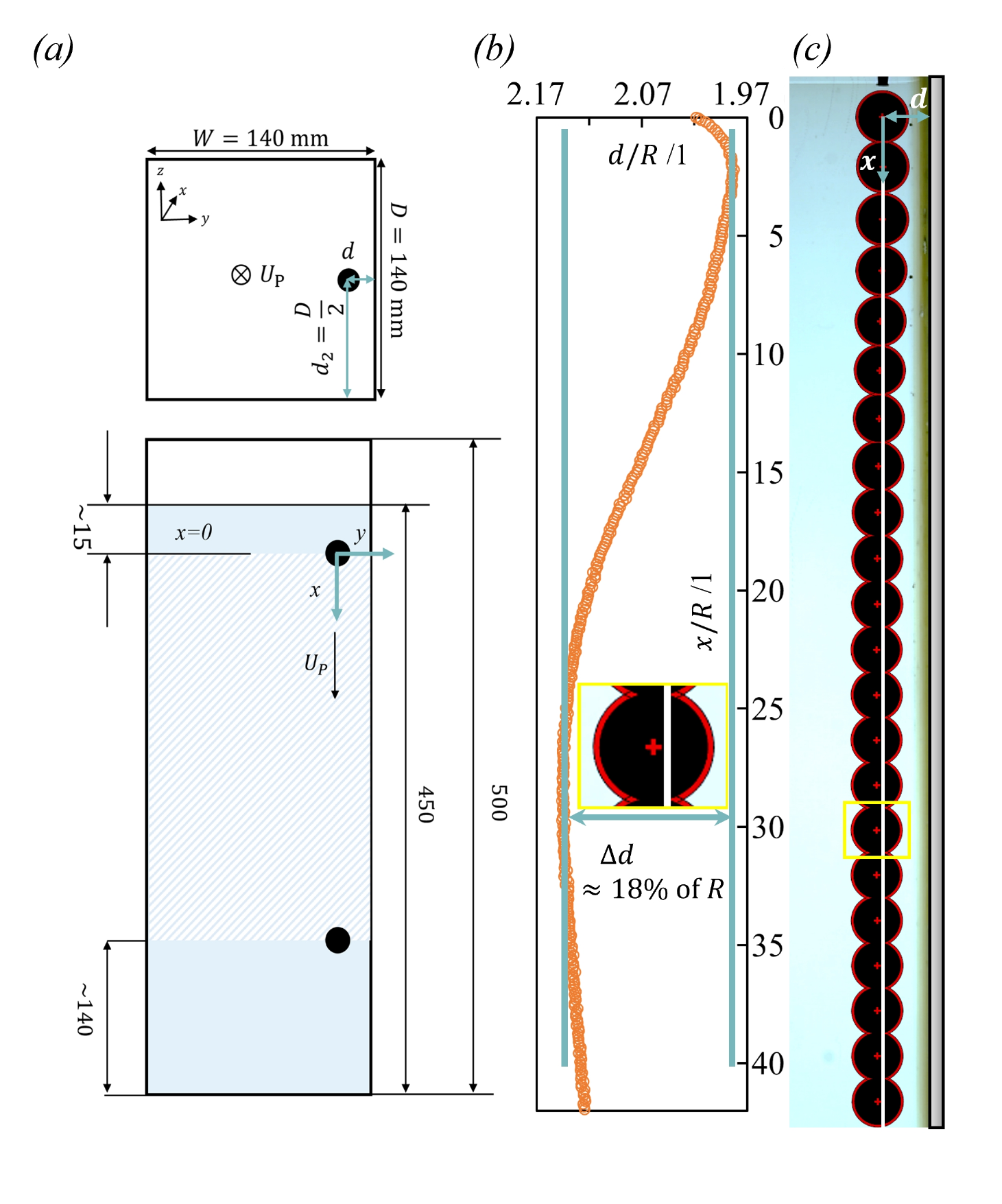}
    \caption{\textit{(a)} Container dimensions in mm (top view and front view); Dashed area: measurement range of experiments near the rigid (glass) wall. The inclination in relation to the vertical (direction of $\textbf{\textit{g}}$) of the right glass wall is $\varepsilon\lessapprox0.2\degree$ (widened downwards); Principle of particle tracking: \textit{(b)} Example of a trajectory represented as $\left(d/R\right) - \left(x/R\right)$-plot; \textit{(c)} merged images of a sedimenting sphere showing the centers and radii of the sphere calculated at different positions. The white line indicates a vertical through the centroid aligned with the vertical of the image.}
    \label{Fig1}
\end{figure}

Sedimentation experiments of spherical model particles were performed in a rectangular duct. As a vessel for the experiments, a glass container of 140~mm x 140~mm x 500~mm (W x D x H) was used, \textit{cf.} Fig. \ref{Fig1} \textit{(a)}. The container was filled up to a height of around 450~mm with silicone oil. The silicone oil had a nominal viscosity of 1000~cSt (measured value of $\eta=0.9797 \pm 0.012$~Pa$\cdot$s using a rotational viscometer IKA\textsuperscript{\textregistered} Rotavisc lo-vi). The liquid density at room temperature was measured to be $\rho=971.28 \pm 0.134$~kg$\cdot$m\textsuperscript{-3} using an analytical balance (Sartorius\textsuperscript{\textregistered} ENTRIS BCE 224i-1s). The choice of these setup properties allowed to investigate the same conditions as small microparticles would experience in aqueous liquids because $Re_\text{P}\ll1$. The spheres were hold with a pipette under weak vacuum and then immersed in the liquid. The pipette tip was equipped with a soft silicone cap which adapted to the shape of the object to be held. This and the weak vacuum prevented the object from being deformed beforehand. The sphere was positioned at the midplane between two opposing walls, having a distance $d$ to the nearest wall, \textit{cf.} top view in Fig. \ref{Fig1} \textit{(a)}. The vacuum was released by slowly opening a valve and flooding the tube with air. This ensured that the spheres were not pushed away from the pipette by air. The silicone oil penetrated the gap between the sphere and the pipette cap, causing the sphere to detach and begin to fall due to gravity. This procedure ensured an initial sedimentation velocity of $U_\text{P, t=0s}=0$~m$\cdot$s\textsuperscript{-1} (starting from rest) while the fluid was initially quiescent $\textbf{\textit{U}}_\text{f, t=0s}= \begin{pmatrix} 0 & 0 & 0\end{pmatrix}^T$. After immersion and between two experiments, a rest period was inserted to exclude the influence of imposed fluid motion and backflows within the closed tank.\newline
\indent High-resolution videos of the sedimentation experiments were recorded by a DSLR camera (Nikon\textsuperscript{\textregistered} D7200 with Sigma\textsuperscript{\textregistered} 50~mm f1.4 objective lens). During the experiments, the container with silicone oil was illuminated from the opposite side of the camera by a collimated light panel. This ensured a sharp contour of the model particle recorded by the camera sensor. The DSLR camera recorded videos of the experiments with a resolution of 1.920~px ($W_\text{px, vid}$) x 1.080~px ($H_\text{px, vid}$). The videos were recorded with the maximum frame rate of 25 frames per second (fps). To avoid blurring effects or artifacts at the edges of the image (\textit{e.g.}, due to spherical aberration), the camera was carefully aligned to image the sedimentation centrally and within the image plane. The optical plane of the camera was aligned perpendicular to the closest wall of the container, which had a negligible inclination of $\varepsilon\lessapprox0.2\degree$ in relation to the vertical in the direction of gravity. Before starting the experiment, an image with higher resolution (3.368~px ($W_\text{px, start}$) x 6.000~px ($H_\text{px, start}$)) was captured. With this start image, the scaling from pixel values to real-length values was performed in each evaluation by scaling via the pipette diameter (known real-length in the image). This procedure allowed to determine the radius of a sphere with an accuracy of $\approx 1\%$ of its radius (size of a pixel $l_\text{px, start}\approx 6.6\cdot10^{-5}$~m, \textit{cf.} \textit{Supplementary information (SI)} §S2). Subsequently, the trajectories and velocities were evaluated with a self-programmed image processing tool in MATLAB\textsuperscript{\textregistered}. The principle of particle tracking is illustrated in Fig. \ref{Fig1} \textit{(b)} and \textit{(c)}. The center of the spheres (centroid of the pixel disks) and the radii of the best-fit circles around the spheres were calculated by the image processing tool. The optical resolution of the camera system during video recording allowed tracking of the centroid in the accuracy range of about 3.5~\% of the radius (size of a pixel in the video $l_\text{px, vid}\approx 2\cdot10^{-4}$~m). The time step between two video images must therefore be chosen large enough to ensure that the centroid of the disc has moved in a distinguishable way. The detail from optical tracking (yellow box in Fig. \ref{Fig1} \textit{(b)}) shows, that the deviation from a vertical in the image was properly resolved optically. Furthermore, it was checked in each evaluation that the spheres moved within the initial optical plane ($xy$-plane) in order to exclude 3D effects, \textit{e.g.}, by evaluating the black discs pixel area $A_\text{px}$, its perimeter $P_\text{px}$ and calculating the isoperimetric quotient $Q=4\pi A_\text{px}P_\text{px}^{-2}$ over time. Further details on optical image tracking accounting for the robustness of the measurement method can be found in the provided \textit{SI}. Furthermore, the reliability of the measurement technique was proven with measurements of rigid spheres in the center of the tank by comparison with existing literature for rectangular channels, \textit{cf.} \cite{Noichl.2022}.\newline
\subsection{Fabrication of model particles} \label{sec3_2}
The spheres used for the experiments in this study were from the same fabrication batches also used in our previous study in the center of the duct.\cite{Noichl.2022}\newline
\indent Elastic model particles with Young´s elastic moduli of $E\approx1712$~kPa, $E\approx936$~kPa and $E\approx135$~kPa (softest sphere) were fabricated from polydimethylsiloxane (PDMS) mixtures (Sylgard\textsuperscript{\texttrademark} 184 and Sylgard\textsuperscript{\texttrademark} 527, Dow Corning\textsuperscript{\textregistered}) by casting and bonding of hemispheres. Base blends of the PDMS elastomers were prepared according to the manufacturer's specifications (Sylgard\textsuperscript{\texttrademark} 184 silicone oil: curing agent in a 10:1 ratio and Sylgard\textsuperscript{\texttrademark} 527 part A: part B in a 1:1 ratio). Each base blend was colored black with 1~w-\% iron oxide powder and degassed. Hereafter, the base blends were mixed with each other in a 1:1 ratio ($E\approx936$~kPa) and a 1:5 ratio ($E\approx135$~kPa), respectively. Pure Sylgard\textsuperscript{\texttrademark} 184, respectively the again degassed mixtures of Sylgard\textsuperscript{\texttrademark} 184 and Sylgard\textsuperscript{\texttrademark} 527, were poured into a mould for hemispheres. The casted hemispheres were hardened for 12~h at 60~\textcelsius. For at least another 36~h, they were cured at room temperature. The measured Young´s elastic modulus of hardened Sylgard\textsuperscript{\texttrademark} 184 base blend was $E\approx1712$~kPa (for details regarding the measurement of elastic properties, \textit{cf.} \cite{Noichl.2022}). In the next step, two hemispheres were bonded together with a thin film of the corresponding newly produced PDMS compound avoiding ridges and asymmetries. A detailed shape analysis is provided in §S2 in the \textit{SI}. The bonded spheres were cured for at least another 48 h at room temperature. The PDMS spheres had a radius of $R\approx6$~mm. The material properties relevant for the experiments are listed in Tab. \ref{tab1} and Tab. \ref{tab2} in section \ref{sec4_1}. It is important to note that, by changing the mixing ratios to vary the Young´s elastic modulus $E$, the density of the material inevitably changes. The variations in the particle-to-fluid density ratio $\gamma=\rho_\text{P}/\rho$ are small (especially those between the PDMS mixtures ($\gamma\approx1.02-1.07$)). However, this had a significant influence on the hydrodynamics, as will be shown in the next chapter.\newline 
\indent Rigid, not manually deformable, spheres with a radius of $R\approx6$~mm and a larger Young´s modulus of 2.9~GPa (according to manufacturer´s specifications) were fabricated by casting and bonding of hemispheres from epoxy resin (casting resin MS 1000 by Weicon\textsuperscript{\textregistered}). Furthermore, commercially available hard polymer spheres with a radius of $R\approx4$~mm were purchased. The measured density of $\rho_\text{R4mm}=1036.4 \pm 2.7$~kg$\cdot$m\textsuperscript{-3} and the rigidity of the material suggest polystyrene (PS). 
The solid density of the rigid and elastic spheres was determined directly after each sedimentation experiment. This ensured to correctly represent the current state of the density. Changes in material properties due to diffusion of oil into the polymer and resulting swelling effects at the surface are avoided due to the short period of time of the experiments, \textit{cf.} density curve Fig. 2 in §S1 in the \textit{SI}. The solid density was determined using a hydrostatic balance (Sartorius\textsuperscript{\textregistered} ENTRIS BCE 224i-1s analytical balance + density determination kit YDK03 for determination of solid densities). The balance has a readability of 0.1 mg. Density measurements were performed in distilled water. \newline
\section{Results and Discussion} \label{sec4}
This section investigates the sedimentation of particles with varying radii and elastic properties near plane, rigid walls in a rectangular container. The study was originally designed as a parameter study focusing on the influence of particle elasticity on the kinematics, with Young’s modulus $E$ selected as the principal control parameter. This choice reflects the fact that differences in $E$ constitute the most apparent and fundamental distinction between the considered materials, motivating a classification into rigid and soft spheres.\newline
\indent During the parameter variation, changes in $E$ inevitably resulted in small variations of the density ratio $\gamma$ and thus of the particle Reynolds number $Re_\text{P}$, \textit{cf.} tables \ref{tab1} and \ref{tab2}. A posteriori analysis reveals that these small changes in $Re_\text{P}$ dominate the observed sedimentation behavior, exceeding the direct influence of elasticity. Throughout the following presentation, it is therefore important to keep in mind that a lower $E$ inherently also means a lower $Re_\text{P}$. Which of these two influential parameters is significant for the specific observation will only reveal itself upon further specific analysis. For clarity of the denomination of the particles, they will consistently be referred to by their softness.

\subsection[Sedimentation of spheres (\textit{R} approx. 6~mm) with varying density and Young´s elastic modulus at varying initial distance to a plane wall]{Sedimentation of spheres ($R\approx6$~mm) with varying density and Young´s elastic modulus at varying initial distance to a plane wall} \label{sec4_1}
\begin{figure}
    \centering
    \includegraphics[width=\textwidth]{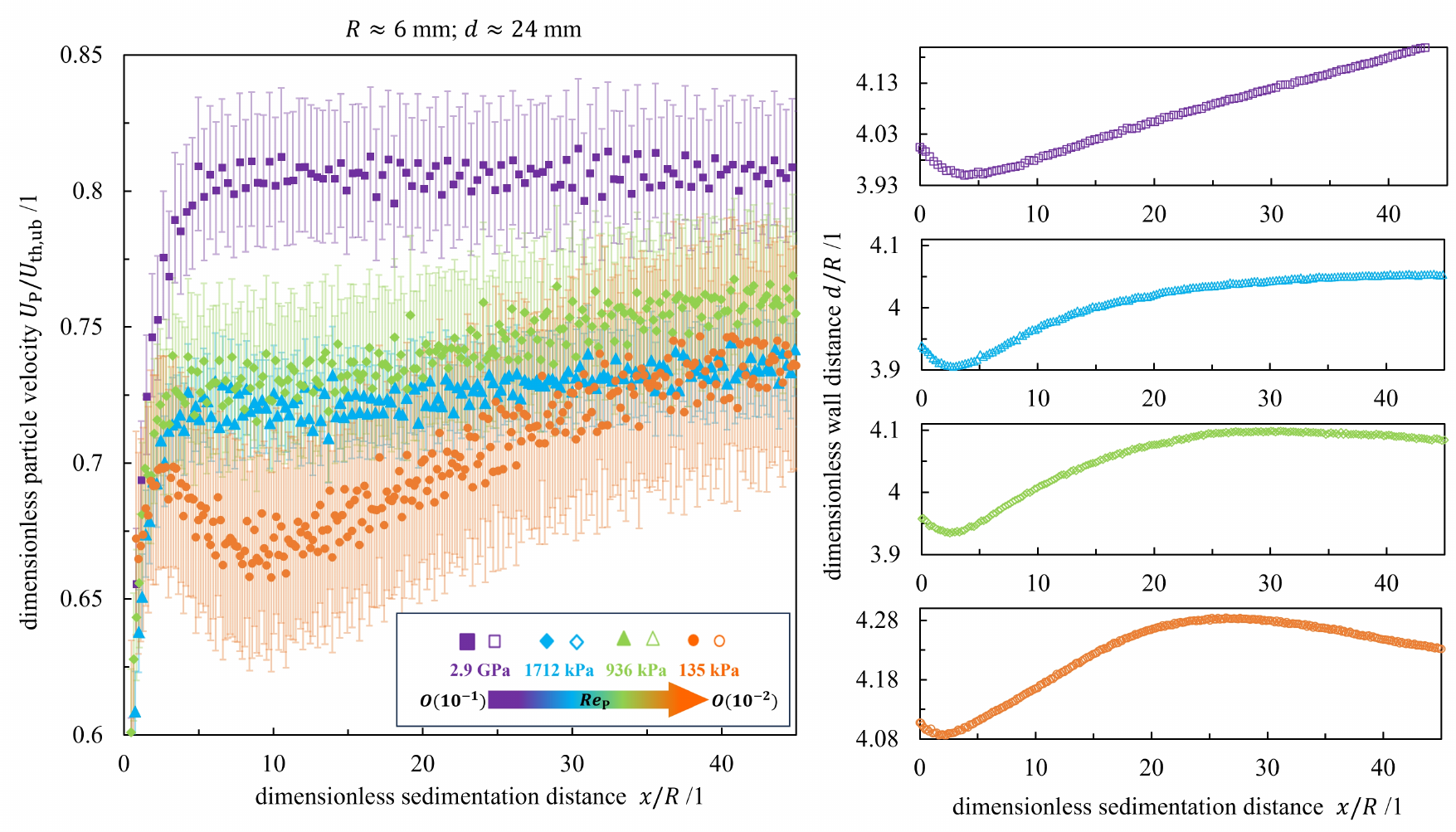}
    \caption{Left: dimensionless particle velocity $U_\text{P}/U_\text{th,ub}$ plotted over the dimensionless sedimentation distance $x/R$ of sedimenting rigid and soft spheres with a radius of $R\approx6$~mm in the vicinity of a plane, rigid wall. The spheres start from an initial dimensionless wall distance of $d/R\approx4$; Right: corresponding trajectories as $\left(x/R\right) - \left(d/R\right)$-plot}
    \label{Fig2}
\end{figure}

\begin{figure}
    \centering
    \includegraphics[width=\textwidth]{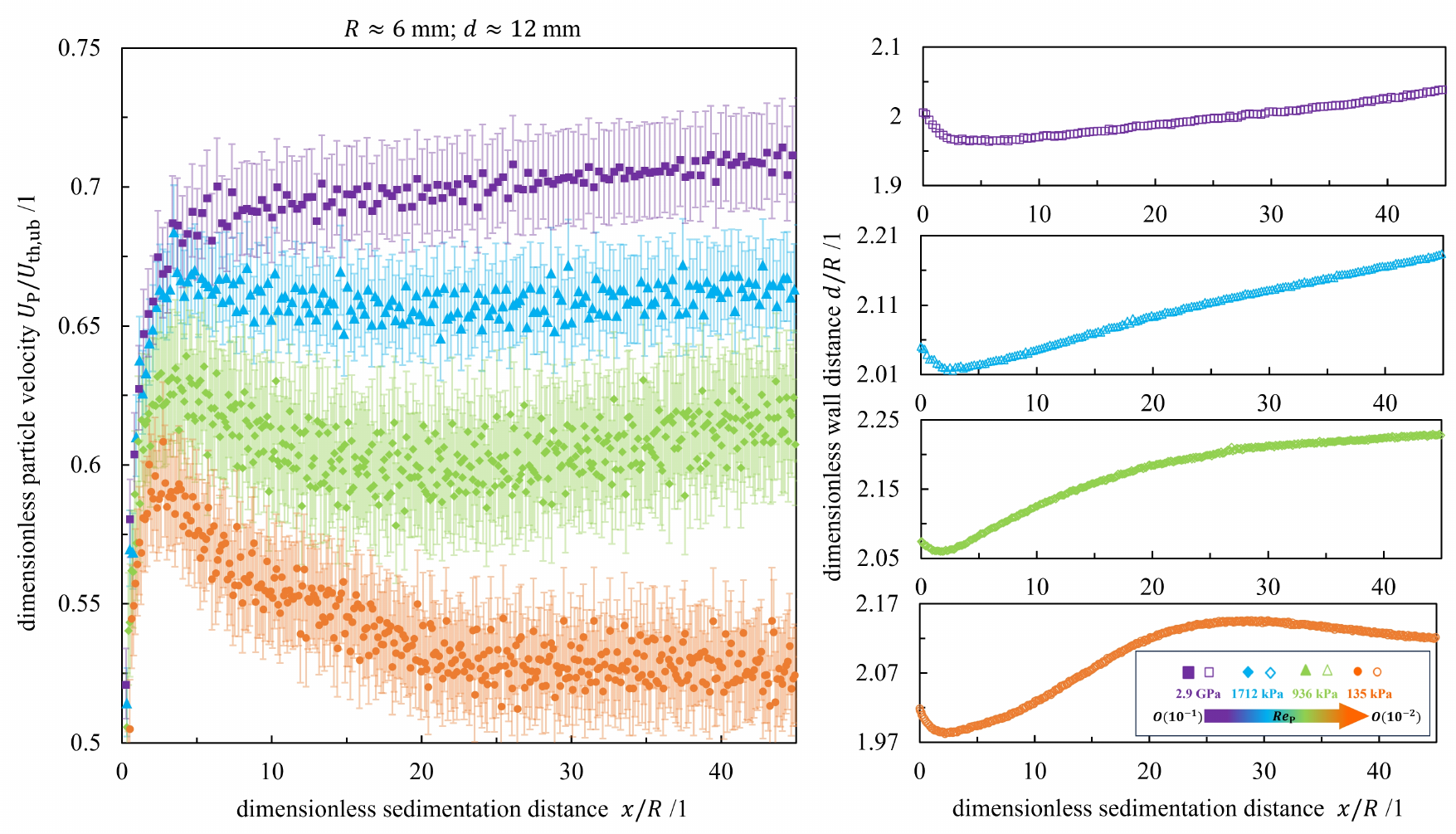}
    \caption{Left: dimensionless particle velocity $U_\text{P}/U_\text{th,ub}$ plotted over the dimensionless sedimentation distance $x/R$ of sedimenting rigid and soft spheres with a radius of $R\approx6$~mm in the vicinity of a plane, rigid wall. The spheres start from an initial dimensionless wall distance of $d/R\approx2$; Right: corresponding trajectories as $\left(x/R\right) - \left(d/R\right)$-plot}
    \label{Fig3}
\end{figure}

In this subsection, the study is focused on the influence of varying elasticity and slightly varying density on the kinematics of sedimenting spheres near rigid walls. All particles had the same size. Experiments were performed at two different distances to the nearest wall: in an intermediate region between the duct center and one of the walls ($d/R\approx4$) and close to one of the walls ($d/R\approx2$). Corresponding results can be found in Fig.~\ref{Fig2} ($d/R\approx4$) and Fig.~\ref{Fig3} ($d/R\approx2$), respectively.\newline
\begin{table}
    \centering
    \renewcommand{\arraystretch}{2.0}
    \begin{tabular}{ c c c c c c c}
        Specimen & $E$ /kPa &  $\rho_{\text{s}}$/ kg$\cdot$m\textsuperscript{-3} & $\overline{\gamma}$& $R$ /mm & $\overline{Re}_\text{Peak}$ & $U_\text{th, ub}$ /mm$\cdot$s\textsuperscript{-1}\\
        \hline
        
        \textcolor{Purple}{$\blacksquare$} \textcolor{Purple}{(rigid)} &	$2.9\cdot 10^6$ & 1160.83 $\pm$ 3.9 & 1.2 & 6.0 & 14.3 $\cdot 10^{-2}$ & 14.9 $\pm$ 0.5 \\
        \textcolor{Cyan}{$\blacktriangle$} \textcolor{Cyan}{(soft)}&	1712 $\pm$ 82.55 &	1036.70 $\pm$ 0.1 &	1.07&	6.2	& 5.0 $\cdot 10^{-2}$ &	5.6 $\pm$ 0.1\\
        \textcolor{YellowGreen}{$\blacklozenge$} \textcolor{YellowGreen}{(soft)} &	936 $\pm$ 33.44&	1007.16 $\pm$ 1.03&	1.04&	6.0&	2.5 $\cdot 10^{-2}$ &	2.9 $\pm$ 0.1\\
        \textcolor{Orange}{\CIRCLE} \textcolor{Orange}{(soft)}&	135 $\pm$ 13.14&	989.40 $\pm$ 0.9 &	1.02&	6.0&	1.2 $\cdot 10^{-2}$ &	1.5 $\pm$ 0.1 \\
    \end{tabular}
        \caption{Physical properties and reference quantities from experiments shown in Fig. \ref{Fig2} (sphere sedimentation experiments with a wall distance of $d/R\approx4$)}
         \label{tab1}
\end{table}

\begin{table}
    \centering
   
    \renewcommand{\arraystretch}{2.0}
    \begin{tabular}{ c c c c c c c}
        Specimen & $E$ /kPa &  $\rho_{\text{s}}$/ kg$\cdot$m\textsuperscript{-3} & $\overline{\gamma}$& $R$ /mm & $\overline{Re}_\text{Peak}$ & $U_\text{th, ub}$ /mm$\cdot$s\textsuperscript{-1}\\
        \hline
        
        \textcolor{Purple}{$\blacksquare$} \textcolor{Purple}{(rigid)} &	$2.9\cdot 10^6$ & $1159.85 \pm 1.51$& 1.2 & 5.9 & 11.4 $\cdot 10^{-2}$ & 14.2 $\pm$ 0.4 \\
        \textcolor{Cyan}{$\blacktriangle$} \textcolor{Cyan}{(soft)}&	1712 $\pm$ 82.55 &	1037.05 $\pm$ 0.25&	1.07&	6.0	& 4.2 $\cdot 10^{-2}$ &	5.3 $\pm$ 0.1\\
        \textcolor{YellowGreen}{$\blacklozenge$} \textcolor{YellowGreen}{(soft)} &	936 $\pm$ 33.44&	1007.16 $\pm$ 1.03&	1.04&	6.0&	2.1 $\cdot 10^{-2}$ &	2.8 $\pm$ 0.1\\
        $\textcolor{Orange}{\CIRCLE}$ \textcolor{Orange}{(soft)}&	135 $\pm$ 13.14&	988.24 $\pm$ 0.39&	1.02&	6.0&	0.9 $\cdot 10^{-2}$ &	1.3 $\pm$ 0.1 \\
    \end{tabular}
    \caption{Physical properties and reference quantities from experiments shown in Fig. \ref{Fig3} (sphere sedimentation experiments with a wall distance of $d/R\approx2$)}
     \label{tab2}
\end{table}

Figures~\ref{Fig2} and \ref{Fig3} present exemplary measurements for the respective experimental configurations (wall distance--Young’s modulus combinations). Each experiment was performed using at least three independent specimens, with at least one repetition per specimen. The observed sedimentation kinematics were reproducible across specimens and repetitions. The measurements are representative within the deviations shown; a complete measurement series for the softest spheres is provided in §S4 of the \textit{SI}. The results are discussed with a focus on a qualitative description of the observed trends.

Individual measurements are shown rather than averaged data. Even unavoidable, small variations in the initial conditions (\textit{e.g.}, the initial wall distance or even slight density variations between specimens of the same material) can have a pronounced influence on the subsequent kinematic evolution under certain conditions that will be shown. Averaging absolute velocity values, trajectories or combining measurements from different specimens would therefore obscure systematic effects and reduce the clarity of characteristic dynamical features. Possible scattering effects were excluded by an appropriate choice of the evaluation time step, which was defined prior to the analysis.

In Figs.~\ref{Fig2} and \ref{Fig3}, the theoretical sedimentation velocity in an unbounded fluid, $U_\text{th,ub}$, is used as a reference for the measured particle velocity $U_\text{P}$. Here, $U_\text{P}$ denotes the velocity magnitude in the $x$-direction, \textit{i.e.}, the direction of gravitational acceleration $\boldsymbol{g}$. This normalization enables a direct qualitative comparison of sedimentation velocities for spheres with different densities and allows the influence of wall proximity to be assessed relative to unbounded sedimentation. The reference velocity $U_\text{th,ub}$ is obtained from a steady-state force balance acting on a sedimenting particle, as described in Eq.~\ref{eq7}.

\begin{equation}
\label{eq7}
6\pi \eta R U_\text{th,ub} \left(1+\frac{3}{16}Re_\text{P,th}\right)=\frac{4}{3}\Delta\rho g \pi R^3
\end{equation}
In Eq. \ref{eq7}, Stokes´s drag is corrected by a factor of $O(Re)$. This correction term was introduced by Carl Wilhelm Oseen in 1910. It accounts for the influence of fluid inertia on the drag force at small but finite Reynolds numbers. Due to the acceleration of mass of the sphere, the velocity field at sufficient distance (the disturbance velocity field) deviates significantly from the velocity field assumed by Stokes. The disturbance increases drag which the fluid exerts on the surface of the sphere.\cite{Oseen.1910} The theoretical particle Reynolds number was calculated by $Re_\text{P,th}=U_\text{th,ub}2R\rho/\eta$. For very small Reynolds numbers, the correction term can be neglected and the theoretical velocity results in the Stokes´s velocity, see Eq. \ref{eq4}. Assessing the Reynolds number in the experiments, $\overline{Re}_\text{Peak}$ in Tab. \ref{tab1} and Tab. \ref{tab2} corresponds to the measured mean Reynolds number shortly after the first transient acceleration phase. As will be discussed, the velocity shows a peak in this area in some cases. $\overline{Re}_\text{Peak}$ is of $O({10}^{-1})$ for the rigid spheres and of $O({10}^{-2})$ for the soft spheres. The assumption that $U_\text{th,ub}$ corresponds to $U_\text{St}$ is valid in good approximation for the soft spheres. However, the influence of inertia on the drag of rigid spheres must be considered. Since Stokes´s drag is smaller than Oseen´s drag, neglecting the inertial correction term for the rigid spheres would result in an overestimation of the theoretical velocity (and consequently in an underestimation of the dimensionless velocity). Therefore, the velocities from experiments with the rigid spheres are nondimensionalized with the theoretical velocity calculated by solving Eq. \ref{eq7} and those of the soft spheres with $U_\text{th,ub}=U_\text{St}$. The error bars shown in the dimensionless velocity curves result from the statistical uncertainties in calculating $U_\text{th,ub}$.

From the purple curves in Fig. \ref{Fig2} and Fig. \ref{Fig3} the following observations for rigid spheres can be made. The velocity curve of the rigid sphere which was released from an initial wall distance in the intermediate region (purple, left diagram in Fig. \ref{Fig2}) is qualitatively comparable with a velocity curve of a rigid, heavy sphere sedimenting in an unbounded fluid. The velocity curve shows one acceleration phase in the very beginning followed by an approximately constant velocity. The sphere accelerated from rest until a dimensionless velocity of $U_\text{P,$\blacksquare$,4}/U_\text{th,ub}\approx0.80$ was reached. The measured velocity was reduced compared to the theoretical value in the unbounded fluid due to the influence of the four surrounding container walls. The value in the intermediate region is slightly lower than the one in the center at $d/R\approx11.7$. For comparison, the dimensionless velocity in the center calculated with $U_\text{th,ub}$ from Eq. \ref{eq7} was measured to be $U_\text{P,$\blacksquare$,11.7}/U_\text{th,ub}\approx0.82$. \cite{Noichl.2022}\footnote{Measurement from section 4.1.1 in \cite{Noichl.2022}: $U_\text{St}$ was used as reference for the velocity of the rigid spheres. For this reason, a smaller value for the dimensionless velocity was reported there.}

Despite the velocity curve appearing classic over the entire sedimentation process, a closer inspection of the trajectory of a rigid sphere sedimenting in the intermediate region reveals unexpected behavior at early times (purple, upper right panel in Fig.~\ref{Fig2}). Immediately after release, the sphere moves towards the wall, \textit{i.e.}, the wall distance initially decreases. This motion persists over the initial acceleration phase until a minimum wall distance is reached. This regime is referred to in the following as \textbf{\textit{inertial wall attraction}} (IWA). The IWA was observed in all experiments near a rigid plane wall, including those with softer and smaller spheres, and is independent of the initial wall distance (\textit{cf.} Figs.~\ref{Fig2}, \ref{Fig3} and \ref{Fig5} in subsection~\ref{sec4_2}). The end of the IWA occurs for all spheres in the range $x/R \approx 2$--$3$.

\begin{figure}
    \centering
    \includegraphics[width=\textwidth]{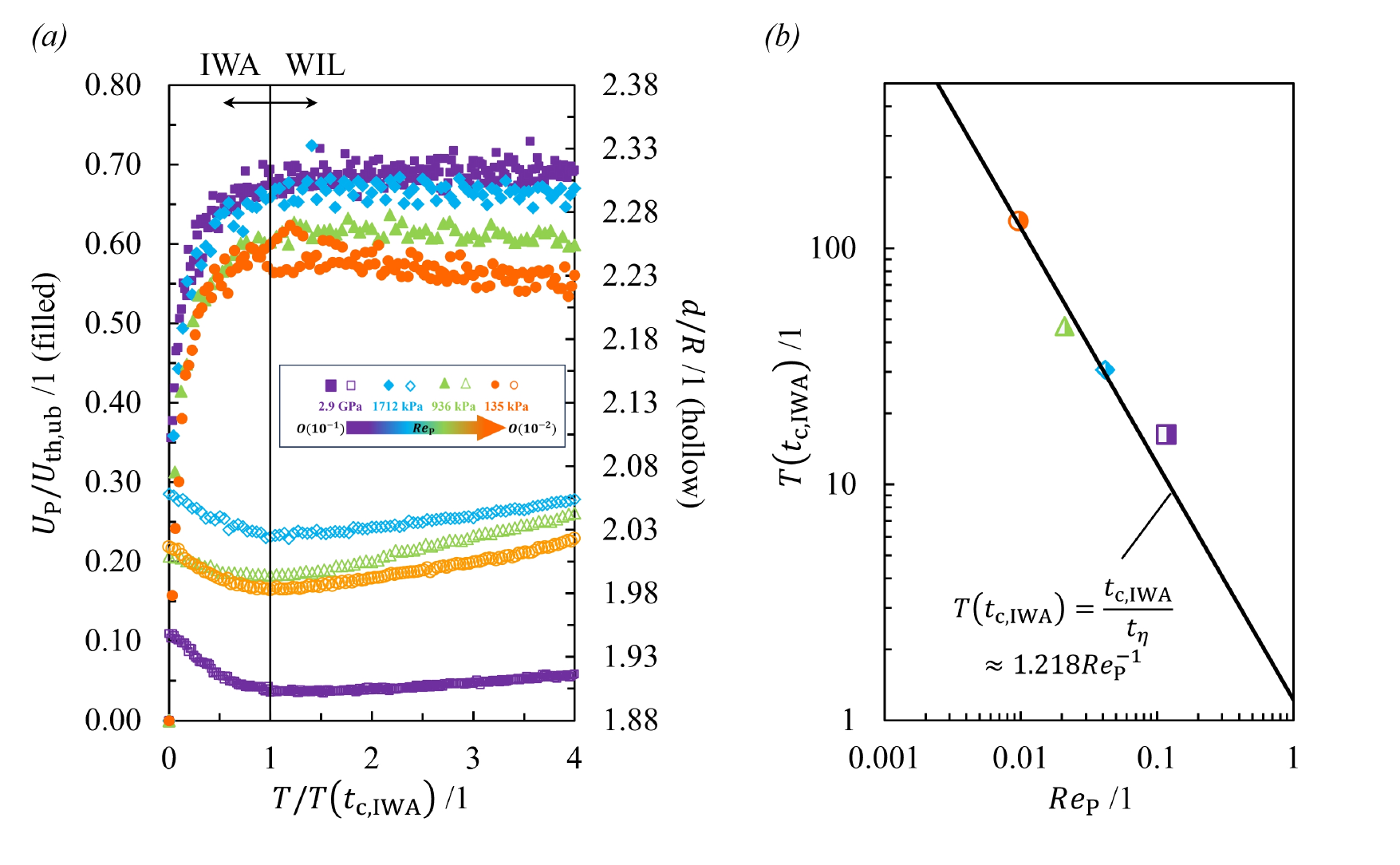}
    \caption{Short-time kinematics with increased temporal resolution for spheres of radius $R \approx 6$~mm released at an initial distance $d/R \approx 2$ (\textit{cf.} experiments shown in Fig.~\ref{Fig3}). \textit{(a)} Dimensionless particle velocity $U_\text{P}/U_\text{th,ub}$ plotted over the dimensionless time $T = t/t_\eta$, normalized by the characteristic time at the end of inertial wall attraction (IWA), $T(t_\text{c,IWA})$. \textit{(b)} Log–log representation of the characteristic time scale $T(t_\text{c,IWA})$ as a function of the particle Reynolds number $Re_\text{P}$, based on the instantaneous particle velocity at $t = t_\text{c,IWA}$, for the spheres shown in Fig.~\ref{Fig3}.}
    \label{Fig4}
\end{figure}

Fig.~\ref{Fig4} shows the short-time kinematics of spheres with radius $R=6$~mm released at an initial distance of $d/R=2$, resolved with increased temporal resolution during the acceleration phase. The dimensionless velocity $U_\text{P}/U_\text{th,ub}$ and wall distance $d/R$ are plotted over the dimensionless time $T = t/t_\eta$, where $t_\eta = 4R^2\rho/\eta$ denotes the viscous diffusion time. The time axis is additionally normalized by the characteristic time at the end of the IWA, $T(t_\text{c,IWA})$ (\textit{cf.} Fig.~\ref{Fig4}\textit{(a)}). Figure~\ref{Fig4}\textit{(b)} shows the corresponding characteristic time scale $T_\text{c,IWA}$ as a function of the particle Reynolds number $Re_\text{P}$.

Normalization with respect to $T(t_\text{c,IWA})$ leads to an initial collapse of the velocity curves, indicating a comparable temporal evolution during the early acceleration phase. A similar normalization approach was proposed by Moorman (1955), who showed that, for spheres in an unbounded fluid, velocity curves at different particle Reynolds numbers can be mapped onto each other using a characteristic time $T_{0.6}$, defined by the time at which $60\,\%$ of the terminal velocity is reached. \cite{Moorman.1955} In the present case, the curves begin to diverge significantly for $T/T(t_\text{c,IWA}) > 1$, indicating a transition to distinct dynamical behavior.

The characteristic time scale associated with the end of the IWA follows approximately an reciprocal dependence on the particle Reynolds number, $T_\text{c,IWA} \propto Re_\text{P}^{-1}$, as shown in Fig.~\ref{Fig4}~\textit{(b)}. Accordingly, the duration of the IWA can be estimated as $t_\text{c,IWA} \approx 1.218\, t_\eta\, Re_\text{P}^{-1}$.

These observations demonstrate that the IWA is a transient, inertia-driven kinematic phenomenon with a well-defined crossover time during the acceleration phase. To the best of our knowledge, such behavior has not been reported previously.

After this first acceleration phase, the rigid sphere sedimenting in the intermediate region showed approximately linear migration away from the wall with $x\sim d$ during the complete experiment, \textit{cf.} purple trajectories in Fig. \ref{Fig2} and Fig. \ref{Fig3}. This is due to the wall-induced lift (WIL) force, $\textbf{\textit{F}}_\text{L}^\text{wi}$, which acts perpendicularly to the wall.\cite{Ekanayake.2021}, \cite{Shi.2020} Such migration away from the wall was shown experimentally by Vasseur and Cox in 1977.\cite{Vasseur.1977b} Vasseur and Cox found that a sphere between two plane parallel walls migrated away from the closer wall until an equilibrium position mid-way between the plates is reached. The alignment of sedimenting particles in the centerline between boundaries was shown numerically, too.\cite{Ghosh.2015} The large, rigid spheres in the experiments shown here did not reach an equilibrium position (independent from the release position). This suggests that migration was not yet finished during the measurement, \textit{i.e.}, the container is too short for the sphere to reach the centerline.\newline 
\indent All in all, one can say that, in the intermediate sedimentation region, the wall-induced lift only affects the trajectory in a noticeable manner but not the velocity. A more detailed curve analysis and further comparisons with theoretical values can be found in §I-a) in Appendix~\ref{secA1_1}.\newline
\indent In contrast to sedimentation in the intermediate region, the effect of the wall-induced lift force $\textbf{\textit{F}}_\text{L}^\text{wi}$ on both the dimensionless velocity and the trajectory is obvious for a sedimenting rigid sphere in the close vicinity to a wall (Fig. \ref{Fig3}, purple curves). In the beginning, there is again an inertial wall attraction during the initial acceleration phase. Thereafter, the velocity keeps increasing. This is due to the decreased drag when increasing the distance to the wall.\newline 
\indent The theoretical dimensionless velocity of a sphere sedimenting near a plane wall at $d/R\approx2$ is $\left.\left(U_\text{P}/U_\text{th,ub}\right)_\text{FO, $\bullet|$}\right|_{d/R=2}=0.72$ (Faxén´s equation with additional Oseen terms, \textit{cf.} Eq. \ref{eqA1} in the Appendix~\ref{secA1_1}). This value is in the range of the measured value at this distance.\newline
\indent All in all, sedimentation experiments of rigid spheres show the expected, classic behavior of sedimentation near walls except for the here-identified inertial wall attraction in the very beginning. Furthermore, the measured values reflect the theoretical approximations.\medskip \newline
In contrast to that, the velocity curves and trajectories of the soft/lower $Re_\text{P}$ spheres differ fundamentally from that of the rigid spheres. Both in the intermediate region when starting from $d/R\approx4$ (Fig. \ref{Fig2}) and in the near-wall region when starting from $d/R \approx2$ (Fig. \ref{Fig3}), none of the soft spheres reached a stationary state within the measurement range. The softer the spheres were, \textit{i.e.}, the smaller $Re_\text{P}$ was, and the smaller the distance from the wall was, the more unexpected the velocity curves and trajectories are. The spheres decelerated after a shortened mass acceleration phase reaching a reduced peak in velocity, although the distance to the wall is significantly increased. This completely contradicts the classic ideas of wall-induced hydrodynamic lift and its consequences. Subsequently, this is called the \textbf{\textit{acceleration-deceleration behavior }}.\newline
\indent None of the soft spheres reached the theoretically predicted value calculated with the corresponding wall distances. Furthermore, the velocity curves and trajectories become curvier, \textit{i.e.}, more non-linear, the softer the spheres were (and thus, the smaller $Re_\text{P}$ was).\newline 
\indent A closer look at the curves reveals a number of relevant points and characteristics of the underlying dynamics which are of great importance for further modeling. One example are inflection points at a specific sedimentation distance of $x/R\approx 10$, \textit{cf.} orange curve in Fig. \ref{Fig2}. A detailed curve analysis can be found in the Appendix.\newline
\indent If the velocity curves of the softest sphere type at various wall distances $d$ are compared directly with each other, another interesting fact emerges. Fig. \ref{FigA1} in the Appendix shows a direct comparison of exemplary velocity curves from sedimentation experiments with wall distances of $d/R\approx2$, $d/R\approx4$ and $d/R\approx11.7$ (center; from previous report \cite{Noichl.2022}) in one figure. It is shown that the curve in the intermediate region can be composed from parts of the other two curves by shifting the curves upwards or downwards (bright curves in Fig. \ref{FigA1}), \textit{i.e.}, the sedimentation velocity in the intermediate region is a mixture of the other two curves. This composition works only for the softest spheres for which $Re\rightarrow0$. This is consistent with the fact that only in this regime the principle of superposition of several hydrodynamic forces can be applied.\cite{Toschi.2019}\newline
\indent This observation raised the question of whether elastic effects, \textit{i.e.} small elastic deformations, could be the main reason for the observed acceleration-deceleration behavior. Or, whether there were additional and perhaps purely hydrodynamic phenomena responsible for the unsteady sedimentation over such large time scales. Therefore, experiments with rigid spheres having particle Reynolds numbers of $O(10^{-2})$ like the soft spheres were performed. By keeping the Reynolds number constant while elastic effects were excluded, the impact of purely hydrodynamic effects should become apparent. \newline
\subsection[Comparison of rigid and soft spheres at similar Reynolds number \textit{Re} approx. \textit{O}(0.01)]{Comparison of rigid and soft spheres at similar Reynolds number $Re_\text{P}\approx O(10^{-2})$} \label{sec4_2}
\begin{figure}
    \centering
    \includegraphics[width=\textwidth]{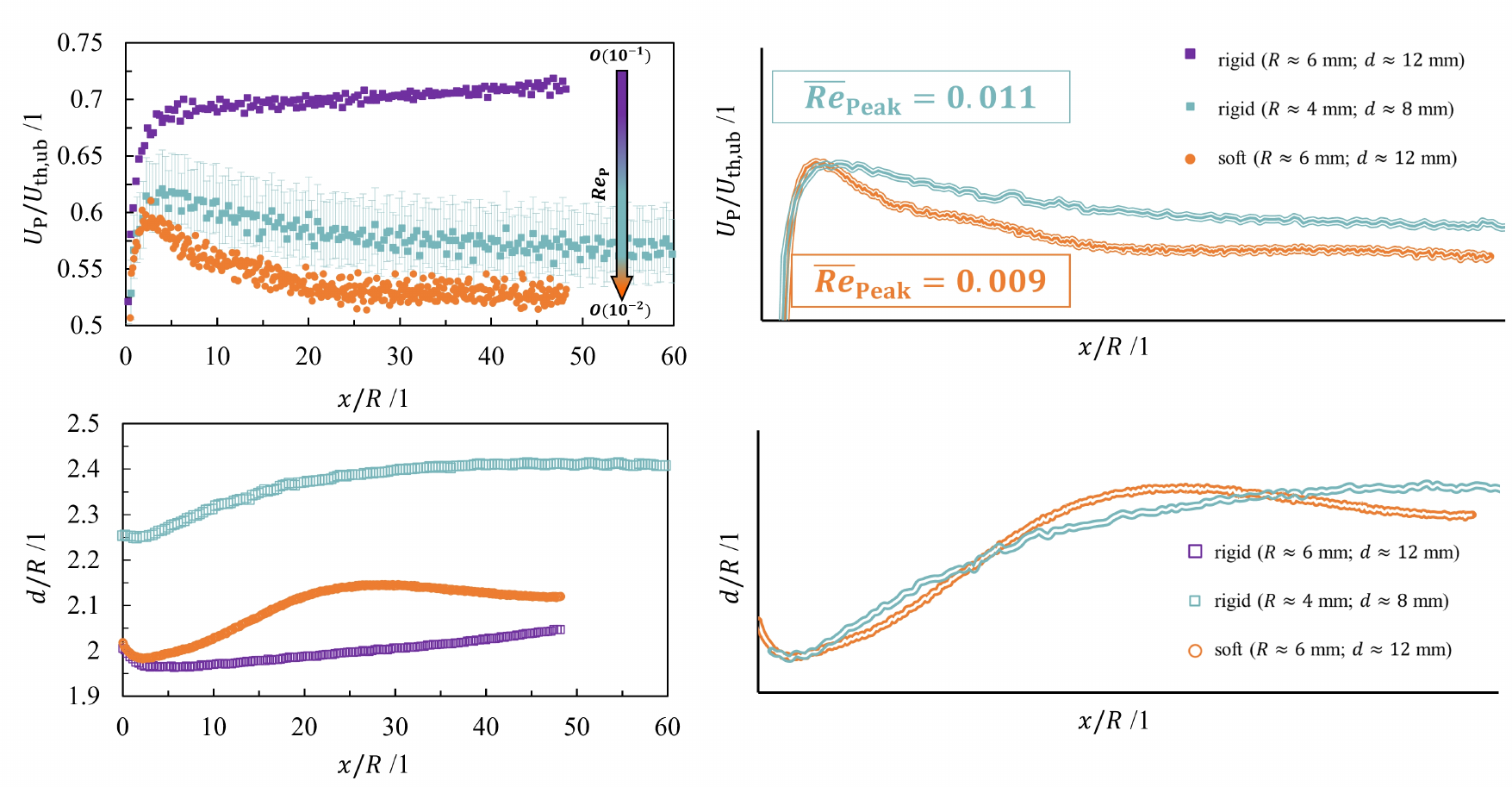}
    \caption{Upper left: dimensionless particle velocity $U_\text{P}/U_\text{th,ub}$ plotted over the dimensionless sedimentation distance $x/R$ of a large rigid sphere sedimenting at $Re_\text{P}\approx O(10^{-1})$ ($R\approx6$~mm, purple), a small rigid sphere sedimenting at $Re_\text{P}\approx O(10^{-2})$ ($R\approx4$~mm, blue-green) and a large soft sphere sedimenting at $Re_\text{P}\approx O(10^{-2})$ ($R\approx6$~mm, orange); Bottom left diagram: corresponding trajectories ($\left(x/R\right) - \left(d/R\right)$-plot); Right figures: qualitative fit curves of the small, rigid sphere and the large, soft sphere sedimenting at $Re_\text{P}\approx O(10^{-2})$ mapped on each other at the point of peak velocity or minimum wall distance, respectively.}
    \label{Fig5}
\end{figure}

Experiments with rigid spheres which had a radius of $R\approx4$~mm and which started at an initial wall distance of $d/R\approx2$ to the nearest wall were performed. The measured peak Reynolds number $\overline{Re}_\text{Peak}$ of these experiments was $O(10^{-2})$. Hydrodynamically, these experiments are thus comparable to the experiments with the softest spheres discussed in the previous subsection. Fig. \ref{Fig5} shows a representative measurement of the velocity (blue-green in the upper left diagram) and a trajectory (blue-green in the bottom left diagram) of a sedimenting rigid sphere with a radius of $R\approx4$~mm released from an initial distance of $d\approx8$~mm (dimensionless wall distance $d/R\approx2$). For better comparability, the curves of the larger rigid sphere ($R\approx6$~mm) and the large soft sphere ($R\approx6$~mm) with a Young´s modulus of $E\approx135$~kPa (softest) from Fig. \ref{Fig3} are shown again. The diagrams on the right show mapped fit curves of the experiments with comparable $\overline{Re}_\text{Peak}$. The corresponding material and size properties (solid densities $\rho_\text{s}$ and radii $R$) and reference quantities are given in Tab. \ref{tab3}. \newline
\begin{table}
    \centering
    \renewcommand{\arraystretch}{2.0}
    \begin{tabular}{ c c c c c c} 
        Specimen &  $\rho_{\text{s}}$/ kg$\cdot$m\textsuperscript{-3} & $\overline{\gamma}$& $R$ /mm & $\overline{Re}_\text{Peak}$ & $U_\text{th, ub}$ /mm$\cdot$s\textsuperscript{-1}\\
        \hline       
        \textcolor{Purple}{$\blacksquare$} \textcolor{Purple}{(rigid)} & $1159.85 \pm 1.51$& 1.2 & 5.9 & 1.14 $\cdot 10^{-1}$ & 14.2 $\pm$ 0.4 \\
        $\textcolor{Orange}{\CIRCLE}$ \textcolor{Orange}{(soft)}&	988.24 $\pm$ 0.39&	1.02&	6.0&	0.9 $\cdot 10^{-2}$ &	1.3 $\pm$ 0.1 \\
        \textcolor{Emerald}{$\blacksquare$} \textcolor{Emerald}{(rigid)} & $1034.5 \pm 2.8$& 1.07 & 4.1 & 1.1 $\cdot 10^{-2}$ & 2.3 $\pm$ 0.1 \\
    \end{tabular}
     \caption{Physical properties and reference quantities from experiments shown in Fig. \ref{Fig5}}
    \label{tab3}
\end{table}
A direct comparison of experiments with rigid spheres in the near-wall region at different Reynolds numbers (a decrease from $Re_\text{P}\approx O(10^{-1})$ to $Re_\text{P}\approx O(10^{-2})$) shows that the acceleration-deceleration behavior also occurs with rigid spheres at $Re_\text{P}\approx O(10^{-2})$. At the same time, the rigid and the soft spheres having similar Reynolds numbers show comparable sedimentation behavior. While the general trend is similar, the soft sphere sedimenting at $Re_\text{P}\approx O(10^{-2})$ shows some superimposed undulations in the trajectory. Consequently, the acceleration-deceleration behavior is not an elastic effect and some other effect seems to be the primary reason for the overall unsteadiness.\newline
\indent Furthermore, based on the detailed curve analysis (see Appendix~\ref{secA1_2}), it is concluded that the undulations in the particle kinematics of the soft sphere during the following phases were induced by elasticity. However, the elastic interaction seems only to be of secondary importance and appears to superimpose the primary effects which seem to be of purely wall-induced, hydrodynamical nature.\newline

\subsection{Computational fluid dynamics simulation of the fluid flow field in a rectangular container} \label{sec4_3}
The assumption of a quiescent fluid within the container is strictly valid only at the initial time, $t = 0$~s. As concluded in the preceding sections, the hydrodynamics, or the flow field developing during sedimentation, respectively, seems to play an important role in the particle kinematics. However, direct experimental access to the flow field, such as velocity-field measurements with particle image velocimetry (PIV), is difficult in the present setup due to the high viscosity of the fluid.

To obtain qualitative insight into the underlying hydrodynamics, a computational fluid dynamics (CFD) simulation was performed. The simulation is intended to analyze an idealized and simplified model system rather than to reproduce the experimental configuration quantitatively. Specifically, the sedimentation of a rigid sphere with radius $R = 6$~mm in a rectangular fluid domain corresponding to the container geometry was considered. The simulations were carried out using COMSOL Multiphysics\textsuperscript{\textregistered} and are based on solving the coupled equations governing the motion of the fluid and the particle.

The fluid velocity field in the spatial frame of the container, $\textbf{\textit{u}}_\text{f}$, is obtained from the time-dependent Stokes (creeping-flow) equations for an incompressible, Newtonian fluid. The conservation of momentum reads
\begin{equation}
    \label{eq8}
\rho \frac{\partial \textbf{\textit{u}}_\text{f}}{\partial t}
= \nabla \cdot \left[-p\textbf{\textit{I}} + \eta \left(\nabla \textbf{\textit{u}}_\text{f} + \left(\nabla \textbf{\textit{u}}_\text{f}\right)^T \right)\right]
+ \textbf{\textit{f}}_\text{ext},
\end{equation}
and the continuity equation reads
\begin{equation}
    \label{eq9}
\rho\nabla \cdot \textbf{\textit{u}}_\text{f} = 0.
\end{equation}
Here, $\textbf{\textit{f}}_\text{ext}$ denotes the external volume force density, given by
\begin{equation}
    \label{eq10}
\textbf{\textit{f}}_\text{ext} = -\rho \left( \frac{\mathrm{d}\textbf{\textit{U}}_\text{P}}{\mathrm{d}t} + \textbf{\textit{g}} \right).
\end{equation}
In this formulation, the fluid is driven by the acceleration of the sphere, such that the particle acceleration $\mathrm{d}\textbf{\textit{U}}_\text{P}/\mathrm{d}t$ provides the coupling between the particle and fluid dynamics.

The particle motion in the direction of gravity $\textbf{\textit{g}}$ is obtained from a force balance,
$\textbf{\textit{F}}_\text{I}^P = \textbf{\textit{F}}_\text{G} + \textbf{\textit{F}}_\text{x,tot}$, leading to the ordinary differential equation
\begin{equation}
    \label{eq11}
m_\text{P} \frac{\mathrm{d}\textbf{\textit{U}}_\text{P}}{\mathrm{d}t}
= m_\text{P} \textbf{\textit{g}} + 2 \int_S \tau_x \, \mathrm{d}S.
\end{equation}
Here, $\tau_x$ is the component of the total traction vector in the direction of gravity, and $S$ denotes the surface of the simulated hemisphere. The integral term represents the total hydrodynamic force exerted by the fluid on the particle. In the steady state, this term corresponds to the sum of buoyancy and drag forces, while in the transient regime it includes all unsteady contributions arising from the evolving flow field. In COMSOL, this term is implemented using the built-in operation $\texttt{intop(-spf.T\_stressx)}$. 

Fluid inertia associated with convective transport is neglected in the governing equations for $\textbf{\textit{u}}_\text{f}$. On the scale of the container, the flow remains close to quiescent in most regions, such that convective effects are small compared with viscous stresses. This approximation is appropriate for capturing the global flow structure, although it does not imply that fluid inertial effects are negligible with respect to the coupled particle kinematics when fully resolved.

\begin{figure}
    \centering
    \includegraphics[width=0.65\textwidth]{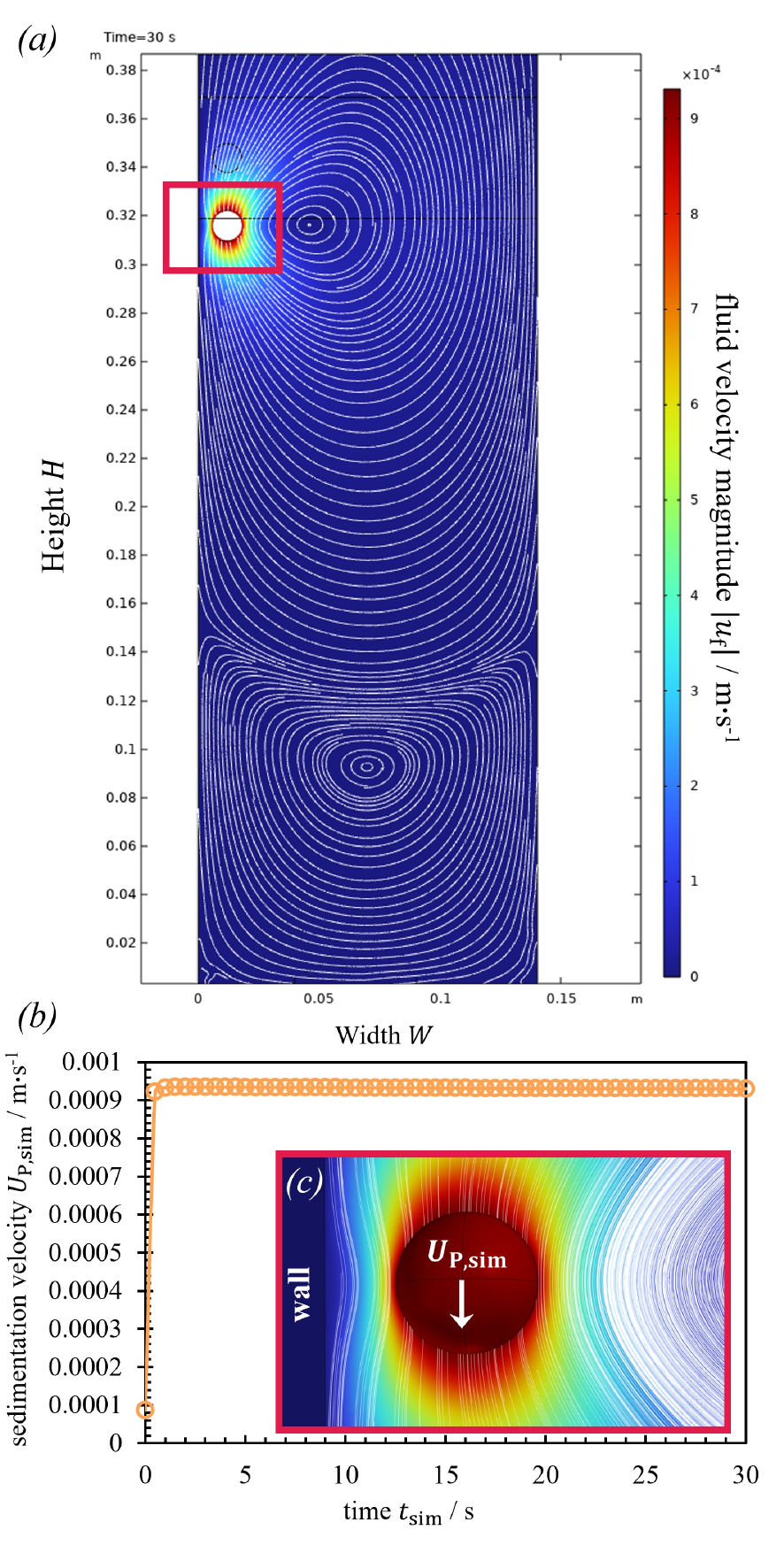}
    \caption{3D simulation of a rigid sphere ($R=6$~mm; $\rho_{\text{s}}=987.59$~kg$\cdot$m\textsuperscript{-3}) sedimenting at $Re_\text{P}\approx O(10^{-2})$ at a distance of $d=12$~mm from a no-slip boundary (magnitude of fluid velocity as contour plot and velocity streamlines on the symmetry plane (half plane trough the container)). \textit{(a)} 2D Streamline plot of the entire simulation domain after $t_\text{sim}=30$~s; \textit{(b)} simulated sedimentation velocity plotted over the simulation time; \textit{(c)} 3D streamline plot corresponding to a close-up of \textit{(a)} }
    \label{Fig6}
\end{figure}
\indent The simulated domain had the same cross-sectional dimensions as the glass container used in the experiments. The measured mean values from the experiments were used as physical properties of the fluid. The solid densities used to solve Eq. \ref{eq11} were $\rho_\text{P,sim}=987.59$~kg$\cdot$m\textsuperscript{-3} (corresponds to experiments with the softest spheres sedimenting at $Re_\text{P}\approx O(10^{-2})$) and $\rho_\text{P,sim}=1159.8$~kg$\cdot$m\textsuperscript{-3} (corresponds to experiments with rigid spheres sedimenting at $Re_\text{P}\approx O(10^{-1})$), respectively.\newline
\indent Simulations were carried out to model sedimentation in the center of the container as well as sedimentation in the near-wall region. Applying appropriate symmetry conditions, the computation of an eighth of the domain was sufficient as a model for the sedimentation in the center, \textit{cf.} Fig. 7 in §S3 the \textit{SI}. In contrast, the 3D domain of the model for sedimentation in the near-wall region ($d_\text{sim}/R=2$) could only be reduced by half, which resulted in a high computational effort. In order to fulfill the symmetry conditions for the entire simulation and to keep the mesh parameters around the sphere surface constant, the distance between the sphere and the boundaries of the simulation domain had to be fixed. Therefore, no spatial displacement towards or away from the wall due to lift forces could be simulated with this model. The center of the sphere had a constant distance to the no-slip boundary of $d_\text{sim}=12$~mm for the near-wall model and $d_\text{sim}=70$~mm, respectively, for the center model. The immersion depth, \textit{i.e.}, the distance of the center of the sphere to the liquid-gas interface, was $h_\text{I,1\&2}=56$~mm (simulation 1 and 2) and $h_\text{I,3}=156$~mm (simulation 3), respectively. The liquid-gas interface was modeled as open boundary with zero normal stress. The sphere was moving in the fluid domain using a moving mesh approach, \textit{cf.} Fig. 8 in §S3 in the \textit{SI}. The sphere was not modeled as a solid body but as a hollow space having the mass of the sphere. This means, that fluid-structural mechanics interactions such as deformations were explicitly not taken into account in this simulation, as the focus of the simulations was on the flow field produced by the fluid displacement of a spherical object. 

It is emphasized that the simulations are not intended as a fully resolved numerical replica of the experimental system in all physical aspects. The model already resolves the full spatial extent of the container together with the fine temporal scales required to capture the slow sedimentation process. This leads to substantial computational demands within the finite element framework. The inclusion of additional physical effects would therefore result in prohibitive computational costs. The simulations consequently address a deliberately simplified, yet well-defined, model problem. The corresponding modeling assumptions---namely the neglect of fluid inertia or the absence of fluid--structure interaction effects, and the exclusion of lateral migration (which would require additional moving meshes)---define the scope of the numerical analysis and must be borne in mind when interpreting the simulation results presented in the following.

The simulated particle velocities over time showed in general a short transient acceleration phase which almost instantaneously transitioned into a stationary state, \textit{cf.} Fig. \ref{Fig6} \textit{(b)}. Although unsteady hydrodynamic forces are inherently contained in the stress-based formulation, the simplified modeling assumptions—particularly the neglect of convective inertia and the idealized boundary conditions—as well as numerical uncertainties in resolving the early transient, may limit the ability of the simulations to reproduce the extended transient behavior observed in the experiments. 

The simulated velocities of the spheres with higher density were slightly larger than the measured velocities ($\Delta_\text{sim;d=70;1159.21}=+5.9$ \% in the center and $\Delta_\text{sim;d=12;1159.8}=+8.3$ \% in the near-wall region). The simulated particle velocity of the sphere with lower density ($\rho_{\text{s}}\approx 988$~kg$\cdot$m\textsuperscript{-3}) sedimenting in the center was in good agreement with the measured velocity at the first plateau (the simulated dimensionless velocity in the center was $\left(U_\text{P,sim}/U_\text{St}\right)_{d=70; 987.59}\approx 0.84$ and the mean measured value of the first plateau $\left(U_\text{P,exp,1st}/U_\text{St}\right)_{d=70; 987.59}\approx 0.83$, \textit{cf.} section 4.2 in the previous report \cite{Noichl.2022}. An overview of all measured and simulated values is shown in Tab. \ref{tabB1} in Appendix~\ref{secB}. It can be concluded, that for the steady state represented in the simulations, both experiments and simulations agree reasonably well for the previously mentioned cases. The simulations reflect fluid-flow processes associated with sedimentation in a rectangular duct (apart from the increase in wall distance due to lift forces, which was not implemented). At the same time, the simulations confirm the experimental observation of a short, steady plateau corresponding to the steady velocity of a heavier sphere, as reported previously. \cite{Noichl.2022} \newline
\indent In contrast, the simulated velocity of a sphere with lower density in the near-wall region was significantly larger ($\Delta_\text{sim;d=12;987.59}=+16.5$~\%) than the measured peak velocity ($U_\text{P,exp,Peak}\approx0.61\cdot 1.3$~mm$\cdot$s\textsuperscript{-1}$\approx 0.8$~mm$\cdot$s\textsuperscript{-1}, \textit{cf.} Fig. \ref{Fig3}). Remarkably, the resulting dimensionless velocity from the simulation in the near-wall region ($\left(U_\text{P,sim}/U_\text{St}\right)_{d=12; 987.59}\approx 0.71$) was almost identical to the corresponding experimental and simulated values for the higher-density sphere ($\left(U_\text{P,exp}/U_\text{th,ub}\right)_{d=12; 1159.85}\approx 0.70$ and
$\left(U_\text{P,sim}/U_\text{th,ub}\right)_{d=12; 1159.8}\approx  0.73$). This supports the observation made in the previous section that the mass acceleration of spheres sedimenting at $Re_\text{P}\approx O(10^{-2})$ is interrupted prematurely at $\left(U_\text{P,exp,Peak}/U_\text{St}\right)_{d=12; 988.24}\approx 0.61$. And this is also in line with the fact that deviations between experiments and simulations occur as soon as new effects start to become important. Naturally, the simulations cannot reproduce the complete experimental kinematics, \textit{i.e.}, the early interruption followed by a decrease in velocity. A straightforward reason is that the sphere was fixed at a constant distance from the wall, thereby constraining the evolution of the fluid dynamics compared with the real system. \newline
\indent The 2D contour plot with streamlines in Fig \ref{Fig6} \textit{(a)} (velocity magnitudes plotted on the half plane of the container) illustrates, that the disturbance around a falling sphere is not confined to a small region around the sphere, even at $Re_\text{P}\approx O(10^{-2})$. Rather, the disturbed area is asymmetric, extended and reaches dimensions on the order of the characteristic length scales of the sphere (\textit{cf.} also the disturbance flow field decaying to 10 \% of $U_\text{P,sim}$ shown in Fig. \ref{FigA2_Disturbance}, Appendix~\ref{secB}).

The streamlines reveal the formation of a slow but noticeable background flow between the sphere and the walls. Owing to fluid reflections at the far wall, a large vortex develops, which is convected downward together with the sphere. The size of this vortex depends on the container width and the wall distance, and it extends both above and below the sphere. 

The vortex shape is influenced by the presence of the free surface (liquid–gas interface) when the sphere is released close to it (simulations~1 and~2), resulting in an asymmetric structure. In contrast, a nearly symmetric vortex is observed when the sphere starts deeper in the fluid domain (simulation~3). The close-up in Fig.~\ref{Fig6}\textit{(c)} shows that the far-wall vortex also affects the local disturbance flow field around the sphere, compressing it laterally (\textit{cf.} Fig.~\ref{FigA2_Disturbance} in Appendix~\ref{secB}). 

Vortex formation during sphere sedimentation has been reported previously in numerical two-dimensional studies \cite{Ghosh.2015}. More recently, Jaroslawski \textit{et al.} (2025) observed experimentally the formation of a vortex ring around a sphere falling in a large container at $0.1 \lesssim Re_\text{P} \lesssim 0.76$, where the assumption of unbounded sedimentation is valid. \cite{Jaroslawski.2025} Particle image velocimetry measurements showed that the vortex core drifts laterally away from the sphere during the transient phase. In confined geometries, such as the present configuration, the unsteady evolution of this vortex structure is subject to additional constraints and may give rise to further kinematic effects, such as the inertial wall attraction. 

Already in 1977, Vasseur and Cox showed that the drag on a sphere sedimenting far from a plane wall can be reduced due to a potential flow induced by wall reflections \cite{Vasseur.1977b}.

\begin{figure}
    \centering
    \includegraphics[width=0.99\textwidth]{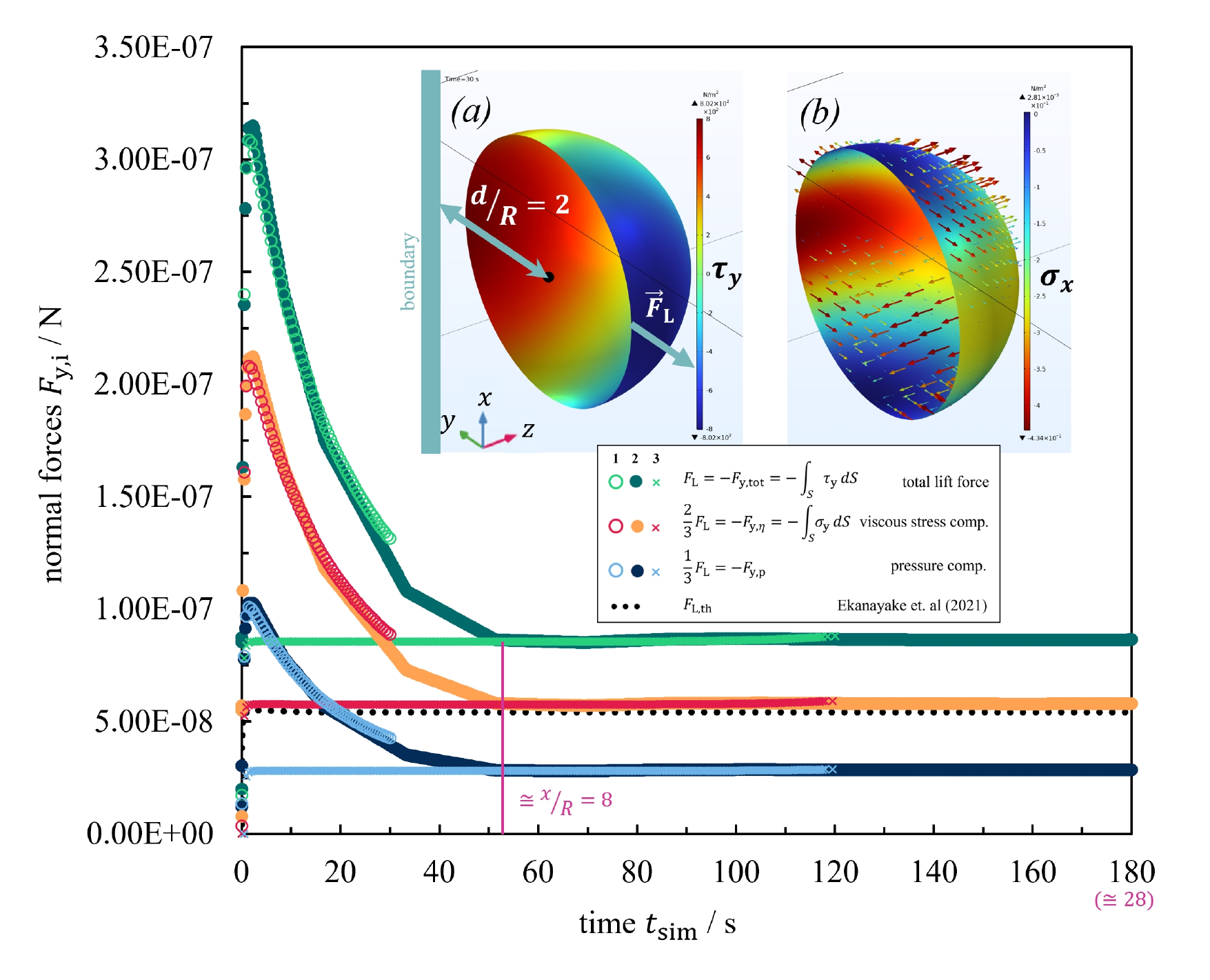}
    \caption{Forces $F_\text{y,i}$ acting on the surface of a sphere in direction normal to the near no-slip boundary (3D simulations of a rigid sphere with a density of $\rho_{\text{s}}=987.59$~kg$\cdot$m\textsuperscript{-3} sedimenting at a distance of $d=12$~mm ($d/R=2$) from the wall). Simulation 1 (hollow circles) and simulation 2 (filled circles) started from an immersion depth of $h_\text{I,1\&2}=56$~mm and from rest. Simulation 3 (crosses) started from an immersion depth of $h_\text{I,3}=156$~mm to the free surface on top. The theoretical value $F_\text{L,th}$ was calculated with the lift coefficient proposed by \cite{Ekanayake.2021}. Insert \textit{(a)}: contour plot of total stresses $\tau_\text{y, sim1}$ on the sphere surface at $t_\text{sim1,max}=30$~s; Insert \textit{(b)}: contour plot of viscous stresses parallel to $\textit{\textbf{g}}$ ($\sigma_\text{x, sim1}$) on the sphere surface at $t_\text{sim1,max}=30$~s. Arrows: viscous stress magnitudes in $y$- and $z$-direction.}
    \label{Fig7}
\end{figure}

 In addition, the simulations illustrate the formation of another vortex on the bottom of the closed container. If the sphere began to fall close to the interface at $H=0.4-h_\text{I,1\&2}=0.344$~m, the vortex extends approximately to a height of $H=0.14$~m, which corresponds to the container width. To exclude effects resulting from this bottom vortex, only the area above this height was considered in the evaluation of the experiments, \textit{cf.} Fig. \ref{Fig1} \textit{(a)} in the methods section.\newline
\indent More interesting estimates can be made by evaluating the fluid stresses and the resulting forces directly on the surface of the sphere, \textit{cf.} the integral term in Eq.~\ref{eq11}. The diagram in Fig. \ref{Fig7} shows the calculated normal forces $F_\text{y,i}$ of a sphere with density $\rho_{\text{s}}=987.59$~kg$\cdot$m\textsuperscript{-3}. The inserts represent plots of the underlying total stresses $\tau_\text{y}$ in direction normal to closest wall (Fig. \ref{Fig7} \textit{(a)}) and the viscous stresses, $\sigma_\text{x}$, at the surface in direction of gravity (Fig. \ref{Fig7} \textit{(b)}). \newline 
\indent The spheres in simulation 1 and 2 started from an immersion depth of $h_\text{I,1\&2}=56$~mm to the open boundary (interface). Simulation 1 (hollow circles) was performed with a higher temporal resolution (\textit{i.e.}, smaller numerical time step which resulted in a high computational effort). The simulation was performed up to a simulation time of $t_\text{sim1,max}=30$~s. In simulation 2 (filled circles) the computational effort was reduced by selecting larger time steps. This reduces accuracy, especially in the transient phase. However, this allowed a total time of $t_\text{sim2,max}=180$~s to be simulated, which is more comparable to the large time scale in the experiment. A simulation time of 180~s corresponds to a dimensionless sedimentation distance of $x/R\approx28$. The modeled sphere in simulation 3 started from an immersion depth of $h_\text{I,3}=156$~mm in order to elucidate the influence of the upper/ top interface.\newline 
\indent In simulations 1 and 2, the total forces in direction perpendicular to the closest wall ($y$-direction), $F_\text{y,tot}$, are strongly time dependent up to $t_\text{sim}\approx 50$~s (corresponding to a sedimentation distance of $x_\text{sim}/R\approx8$). This long timescale, as well as the large distance $x/R$, goes far beyond the initial phase of mass acceleration. The end of the initial mass acceleration (at $t_\text{sim}\approx 2$~s or $x_\text{sim}/R\approx 0.3$) corresponds to the maximum lift force. Since both simulations, \textit{i.e.} simulations with small and large time-steps, show the same time dependence over such long timescales, numerical reasons for this force overshoot are excluded. The maximum magnitude of the force overshoot is more than 3.6 times larger than in the stationary state. Such an overshoot was also observed in the simulation of spheres with a higher density ($Re_\text{P}\approx O(10^{-1})$), but the magnitude was considerably lower with 1.1 times the steady-state value, \textit{i.e.} smaller force gradients were observed.\newline
\indent The simulated total lift force can be compared with values of the total lift force of the model for translational motion parallel to one plane wall which was proposed by Ekanayake \textit{et al.}, \textit{cf.} equations \ref{eq3} and \ref{eq4}. \cite{Ekanayake.2021} Their model, also based on numerical simulations, employed a sufficiently large domain such that distant walls did not influence the forces (geometry and mesh independence was achieved). The theoretical value shown in Fig. \ref{Fig7} was calculated with the simulated velocity $U_\text{P,sim}$, and the corresponding lift force is instantaneously steady. Consequently, the strong time dependence of the lift force at the beginning cannot be mapped with the existing model nor can it arise from fluctuations in the particle velocity, which originate, for example, from numerical inaccuracies. Apart from the initial time dependence, the stationary values are of the same order of magnitude, with the steady lift force in a rectangular duct being larger compared to the lift force induced by a single plane wall (the apparent correspondence with the viscous component of our simulations is coincidental).

Remarkably, corresponding dynamical crossover points were also identified in the experiments at $x/R \approx 10$. At comparable length and time scales, particles sedimenting at lower particle Reynolds numbers ($Re_\text{P} \approx O(10^{-2})$) exhibit a similar crossover in their kinematics, manifested by a change from deceleration to acceleration; see, for example, the orange curve in Fig.~\ref{Fig2} and the detailed curve analysis in §II-a) of Appendix~\ref{secA1_1}.

In contrast, the normal forces obtained from simulations in which the sphere was initially deeply immersed (simulation~3, crosses) do not exhibit such time dependence. This indicates that proximity to the free interface has a significant influence on the dynamics. The pronounced force gradients are likely a direct consequence of vortex formation and its initial asymmetry, caused by the interaction of the far-wall vortex with the interface. 

In the experiments, the immersion depth was smaller ($h_\text{I,exp}\approx 15$~mm), \textit{cf.} Fig.~\ref{Fig1}. The influence of the interface on the evolving vortex structure in the container may therefore have been even more pronounced and could contribute to the unsteady phenomena observed at early times, \textit{i.e.} the \textit{inertial wall attraction}. Due to design constraints, larger immersion depths in the near-wall region could not be realized in the present experimental setup and will be addressed in future investigations.

Alternatively, one can think of this first transient phase in the following manner: the total force on the sphere is the sum of the stresses around the sphere. The flow around the sphere can be thought of consisting of the flow between the sphere and the closest wall and the flow on the other side facing the larger part of the container. Immediately after release, the near-wall flow field develops almost instantaneously and is characterized by the wall distance as the relevant length scale. Its contribution to the total force is therefore almost constant. In contrast, the flow field on the opposite site requires much longer to develop and depends on the size of the vortex, \textit{i.e.}, in the present case, the size of the container. Consequently, its contribution to the total force evolves until the vortex is sufficiently developed. It is therefore clear, that the total force will initially vary. Remarkably, the observed distance $x_\text{sim}/R\approx8$, during which the force varied, corresponds to the size of the vortex, \textit{i.e.}, the distance between the sphere and the opposite wall. A similar behavior would also be expected in other geometries, \textit{e.g.} if the upper boundary were solid. In that case, the vortex would likewise need to develop into a sufficiently symmetric shape before the force became stationary. \newline
\indent Finally, another quantity that can be estimated by the simulations is the local deformation caused by the shear stresses. The maximum total fluid stresses, or total traction, respectively, in the $y$ and $z$ direction are located in the gap between the sphere and the wall and are $|\tau_\text{y;z}|_\text{max}\approx 800$~Pa. Assuming linear elasticity, this corresponds to a linear deformation of $\Delta R_\text{y;z}=|\tau_\text{y;z}|_\text{max} E^{-1} R\approx 35$~µm. The approximation of the deformation with the purely viscous component in direction of gravity ($|\sigma_\text{x}|_\text{max}=0.434$~Pa, \textit{cf.} red colored region on the sphere surface in Fig. \ref{Fig7} \textit{(b)}) leads to much smaller theoretical deformations due to shear. Based on the shear modulus $G\approx E/3$, the approximate deformation would be $\Delta  R_\text{x}=3|\sigma_\text{x}|_\text{max} E^{-1} R\approx0.06$~µm. This shear deformation is almost negligible in absolute terms and at first glance it cannot be assumed that these small deformations purely due to viscous shear would have an influence on the kinematics (\textit{cf.} perturbation parameter $\alpha$ from Murata discussed in our previous report).\cite{Murata.1980b}, \cite{Noichl.2022} Nevertheless, it should be mentioned that these small shear deformations are distributed asymmetrically on the surface of the sphere and are largest in the area between the sphere and the wall, see Fig. \ref{Fig7} \textit{(b)}, red area of the contour plot. This should be taken into account with regard to the hydrodynamic interactions and the long time scales on which they can act at $Re_\text{P}\approx O(10^{-2})$.\newline

\section{Summary and Conclusions} \label{sec5}
In this study, the sedimentation of spherical particles starting from rest near a plane wall within a rectangular duct was investigated. Experiments were performed at low particle Reynolds numbers ($Re_\text{P}\lessapprox O(10^{-1})$), and included elastohydrodynamic interactions. A range of previously unreported phenomena was observed, including persistent unsteady motion, nonlinear kinematics, and distinct features such as \textit{inertial wall attraction} and superimposed undulations in the trajectories and velocity curves of soft spheres.

Rigid spheres sedimenting with particle Reynolds numbers of $Re_\text{P}\approx O(10^{-1})$ exhibited kinematics consistent with classic wall-lift and drag models. In contrast, spheres with decreasing Young´s moduli and smaller density, corresponding to smaller particle Reynolds numbers $Re_\text{P}\approx O(10^{-2})$, displayed increasingly atypical behavior. At first glance, both trajectories and particle velocities appeared to depend strongly on the elastic modulus: with increasing elasticity, velocity curves and trajectories became increasingly nonlinear. This behavior was also observed when particles sedimented far from the wall, for example in the intermediate region between the duct centerline and one wall. It was however found that in the most cases, the particle Reynolds number was the more relevant influential factor.

A detailed analysis of the velocity curves and trajectories revealed that parts of the observed motion contradict established theories of hydrodynamic interactions between rigid particles and rigid walls. In particular, existing steady wall-lift models proved insufficient to describe the motion of soft spheres under these conditions. A characteristic \textit{acceleration–deceleration behavior} was observed: initially, the softest spheres accelerated from rest to a velocity below the theoretical settling velocity. After reaching a peak velocity, they abruptly ceased accelerating and decelerated strongly, even while continuing to migrate away from the wall. This behavior contradicts classic wall-lift predictions, which imply an increase in settling velocity with increasing wall distance due to reduced drag. In other cases, the opposite trend was observed, with spheres accelerating while decreasing their wall distance, suggesting an apparent attraction towards the wall.

With respect to such attraction phases, a to date unrecognized kinematic phenomenon was identified: the \textit{inertial wall attraction} during the mass acceleration. Both rigid and elastic particles with Reynolds numbers in the range $10^{-2}\lessapprox Re_\text{P} \lessapprox 10^{-1}$ migrated towards the wall immediately after release from rest, rather than away from it. This inertial wall attraction is a purely hydrodynamic, wall- and interface-induced inertial effect occurring for all particles sedimenting in this hydrodynamic regime under these experimental conditions. A characteristic time scale associated with the end of the inertial wall attraction was identified, which scales inversely with the particle Reynolds number, $t_\text{c,IWA} \propto Re_\text{P}^{-1}$.

Further comparison between elastic and rigid spheres sedimenting at comparable Reynolds numbers showed that purely hydrodynamic effects must play a primary role in the deceleration following the initial acceleration phase. Rigid spheres sedimenting at $Re_\text{P}\approx O(10^{-2})$ also decelerated despite increasing their wall distance, demonstrating that this behavior is not caused by elasticity alone. Unlike the soft spheres, however, deceleration of rigid spheres was nearly linear and less pronounced. The additional nonlinearities observed in the velocity curves and trajectories of soft spheres are therefore attributed to elastic effects, which appear to be superimposed on, and secondary to, the dominant hydrodynamic mechanisms.

To shed light on the flow field and the mechanisms underlying the observed transient behavior, a computational fluid dynamics (CFD) simulation of a sphere moving through a rectangular fluid domain was performed. The simulation was deliberately simplified: fluid–structure interactions in the sense of deformable solid–fluid coupling were not explicitly resolved, and the sphere was modeled as a hollow body with the mass of a solid sphere. The fluid flow was described by the time-dependent Stokes equations, and the coupling between particle motion and the surrounding flow field was realized through a stress-based formulation.

Using a moving-mesh approach, the simulations revealed that lift forces during gravitational acceleration are strongly time dependent in the vicinity of open boundaries. The computed velocity field showed that the disturbance around the sphere extends well beyond its immediate vicinity, reaching length scales comparable to the sphere diameter. In addition, the streamlines indicate the formation of a slow but noticeable background flow between the sphere and the wall due to fluid reflections, as well as the development of a large vortex structure between the sphere and the walls far away. These flow features evolve in time and contribute to the unsteady forces acting on the particle.

Based on the combined experimental and numerical observations, the following mechanisms are proposed for the observed transient phenomena. At early times during mass acceleration, when the particle velocity is still small compared to the peak velocity, ($Re_\text{P, $t\gtrapprox$0 s}\ll Re_\text{Peak}$), the far-wall reflections and the evolving large-scale vortex induce forces that dominate over near-wall effects such as stationary Faxén lift forces. In the experiments, where the spheres were free to rotate and to migrate, these forces due to the vortex background flow may alter the trajectory in such a way that the sphere appears to be attracted. This effect is likely amplified by the proximity to the liquid-gas interface, which breaks the symmetry of the evolving vortex.

These wall-and interface-induced hydrodynamic effects are related to the inertia of the surrounding fluid. In order to clarify the term “inertial forces” in general, the different types of inertia involved in the sedimentation process and their different origins are reviewed.

\textbf{\textit{Inertia due to mass acceleration}}: During acceleration, the inertial forces which are related to the inertia of mass determine the particle dynamics. These are the inertial forces which depend on the unsteady particle velocity ($\text{d}\textbf{\textit{U}}_\text{P}/\text{d}t\neq0$). On the one hand, there is the pure mass inertia $m_\text{P} \frac{\text{d} \textbf{\textit{U}}_\text{P}}{\text{d}t}$ which results from accelerating the mass of the particle from rest. And on the other hand, there are coupled particle-fluid inertial forces, \textit{e.g.} the added mass or the Basset history force during the initial mass acceleration (\textit{cf.} Eq. \ref{eq1}). In the sedimentation process, these inertial forces are decisive for the appearance of the velocity curve during the initial acceleration phase. In general, if particle-fluid inertial forces are important (especially when $\rho_\text{P}/\rho \rightarrow1$), they lead to an extended transient mass acceleration phase compared to heavier particles, \textit{i.e.}, a less steep increase of the velocity curve in the beginning.

\textbf{\textit{Inertia due to fluid advection}}: Another type of inertial forces that can act on a particle are forces due to fluid advection, \textit{i.e.} the \textit{reaction force} to the inertia of the surrounding fluid. Advective inertia can either originate from disturbance flows. Disturbance flows are the flows caused by advection of the fluid volume $V_\text{P}$, which was displaced by the particle, see Faxén´s law in Eq. \ref{eq2}. Or advective inertia on a particle can originate from separately imposed background flows. In general, advective inertial forces are forces which depend on the velocity of the surrounding fluid $\textbf{\textit{U}}_\text{f}$, \textit{cf.} Eq. \ref{eq1}.

\textbf{\textit{Hydrodynamic history effects}}: The most unintuitive and least understood type of inertial forces are the long-time persistent unsteady forces or hydrodynamic history effects, respectively. In this case, “long-time persistent” means time scales that go far beyond the initial mass acceleration. The Basset force is also a history force, but refers specifically to the phase shortly after acceleration from rest, \textit{i.e.}, before mass inertia decays and drag forces become dominant. Hydrodynamic history effects can result, for example, from inhomogeneities in \textit{e.g.} the shape or the material composition, or, \textit{e.g.} from deformations of the surface. The unsteady forces come into play, when such inhomogeneities disturb the surrounding fluid, or fluid flow respectively, and the object sediments so slow, that "it can feel the hydrodynamic history in the fluid produced by itself".

M.V. Díaz (2021) proposed to add, in addition to the Basset history force, a second history force to the force equilibrium of the BBO equation. This additional history force has a non-singular kernel to account for inhomogeneities.\cite{Diaz.2021} The regarded unsteady forces could also play a role, when spherical or non-spherical objects sediment in a time-dependent background flow, like oscillating flows. \cite{Lawrence.1986}, \cite{Lawrence.1988}, \cite{Seyler.2019}, \cite{Ghosh.2015} 

Among other things, Feng and Joseph concluded from their simulations in 1995, that the presence of walls also tends to enhance unsteady inertial effects. In their report they stated: ”In the literature, the practice of completely ignoring the unsteady effects of inertia at low Reynolds numbers is overwhelming.”\cite{Feng.1995} This circumstance of ignorance, intentionally or unintentionally, has not changed much until today. The literature is partly very inconsistent concerning these unsteady effects of inertia. This is certainly also due to the fact that it is difficult to clearly distinguish between the different origins and influences on such unsteady forces, as the three types of inertial forces are coupled in many cases.

Rallabandi (2021) presented a modification of the Maxey–Riley equation (Eq. \ref{eq1}), in which the inertia of a spatially varying background flow appreciable at the particle scale---such as for large vortex radii observed in our experiment---is included. His analysis predicts that inertial forces associated with second gradients of the background flow velocity can amplify inertial Faxén terms by a factor of three, rendering them dominant relative to wall-induced Faxén lift forces. \cite{Rallabandi.2021} These theoretical estimates support the experimental observation of inertial wall attraction and are consistent with the conclusions drawn from the CFD simulations.

For spheres sedimenting in the near-wall region---\textit{i.e.} with the wall located in the inner region of the Stokes length scale $L_\text{S}$, \textit{cf.} \cite{Ekanayake.2020}---it is commonly assumed that, once mass-acceleration effects decay, drag and wall-induced lift forces drive migration away from the wall. However, the present experimental results show that this assumption holds only for particles reaching $Re_\text{Peak}\approx O\left(10^{-1}\right)$. At smaller peak Reynolds numbers, $Re_\text{Peak}\approx O(10^{-2})$, not only fluid inertia due to advection, \textit{i.e.} the evolving background flow, but also the particle–fluid inertial forces, such as the Basset history force and added mass, play a significant role. As a consequence, the sum and coupling of these types of inertial forces during the mass acceleration phase damp the dynamics in this regime further and result in a shift to longer time scales, at which unsteady forces (\textit{cf.} type \textit{Hydrodynamic history effects}) become significant and manifest as instationarities causing, for example, the \textit{acceleration-deceleration behavior}. At larger Reynolds numbers, particles traverse the flow too rapidly to be affected by such history-dependent or wall-reflected disturbances. In the presence of elasticity, the complex coupling between fluid inertia, hydrodynamic history effects and elasticity-induced forces gives rise to additional nonlinearities, which are referred to here as an \textit{elastohydrodynamic memory effect}.

\begin{figure}
    \centering
    \includegraphics[width=\textwidth]{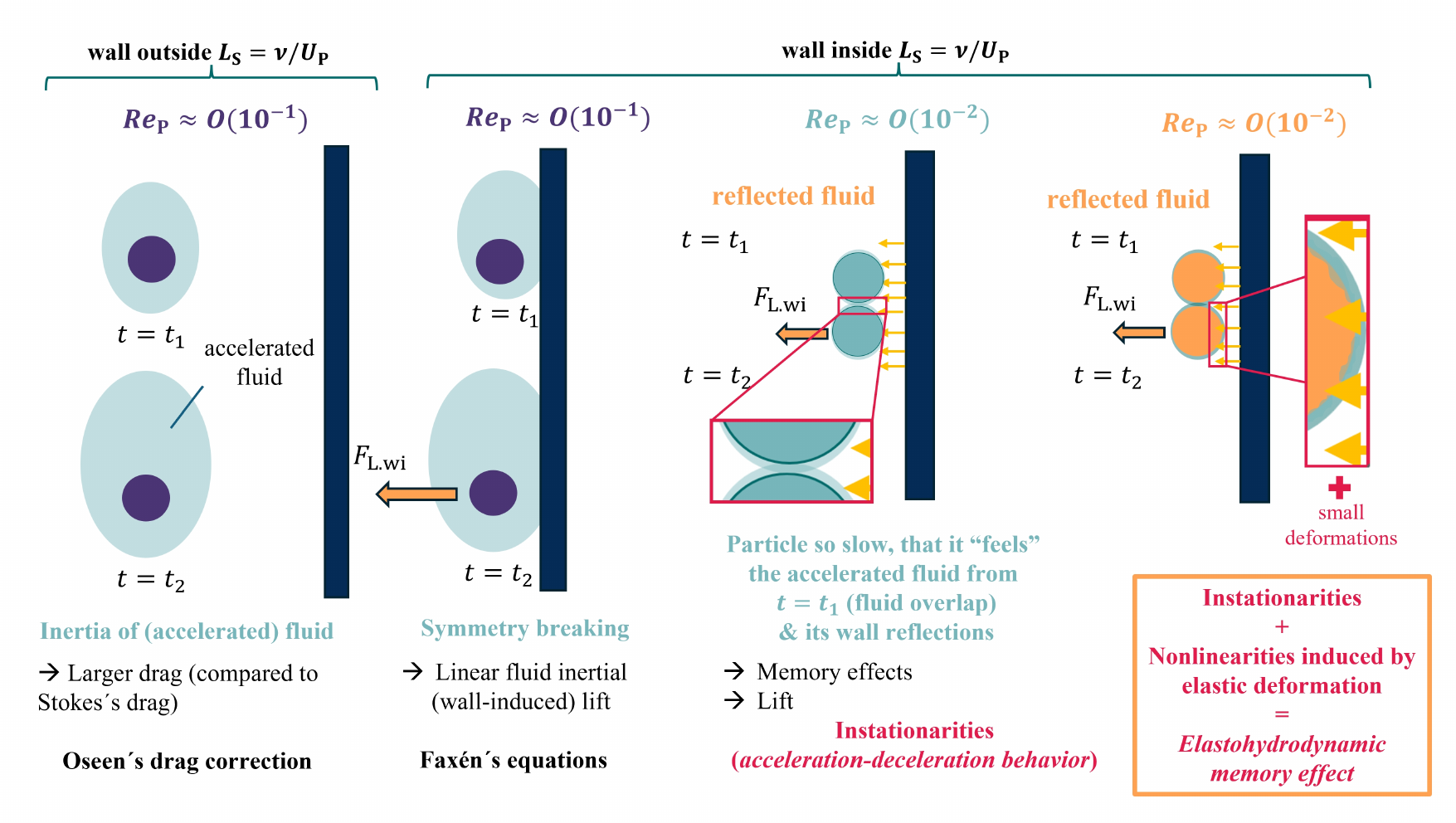}
    \caption{Mechanisms of spherical particles sedimenting at $Re_\text{P}\approx O(10^{-1})$ and at $Re_\text{P}\approx O(10^{-2})$ near a plane wall. Left: sedimentation at $Re_\text{P}\approx O(10^{-1})$ with walls outside the Stokes length scale $L_\text{S}=\nu / U_\text{P}$. Right: sedimentation in the creeping flow regime with walls inside the Stokes length scale $L_\text{S}=\nu / U_\text{P}$. The schematically represented disturbance flow fields (accelerated fluid) are to be regarded as a fluid shell with comparable outer velocity.}
    \label{Fig8}
\end{figure}

Fig.~\ref{Fig8} summarizes the presumed mechanisms governing the sedimentation of spherical particles near walls in the creeping-flow regime ($Re_\text{P}\lessapprox0.1$), spanning approximately one order of magnitude in particle Reynolds number. The figure contrasts spheres of equal size translating near walls located either outside or inside the Stokes length scale $L_\text{S}$ (purple: rigid spheres at $Re_\text{P}\approx O(10^{-1})$; blue–green: rigid spheres at $Re_\text{P}\approx O(10^{-2})$; orange: elastic spheres at $Re_\text{P}\approx O(10^{-2})$).

The disturbance flow field is illustrated schematically as a shell surrounding the particle. Across the considered Reynolds numbers, the shells are drawn with comparable velocities at their outer boundaries and therefore do not represent the physical extent of the disturbance field relative to the particle size. Within this region, fluid inertia remains significant compared with the outer flow. If the shell intersects a wall, the resulting symmetry breaking induces an approximately linear lift force that can be described by Faxén's equations, with Oseen drag corrections at the larger particle Reynolds numbers.

For slower particles at $Re_\text{P}\approx O(10^{-2})$, the region of accelerated fluid is smaller owing to the reduced momentum transfer. At the same time, particle–fluid inertial contributions, such as the Basset history force, shift the dynamics to longer time scales. These particles therefore interact with the wall for longer times and can sense reflected disturbances even when the shell itself does not intersect the wall. In this regime, additional effects may arise, including hydrodynamic history associated with elastic deformation, which may lead to the presumed \textit{elastohydrodynamic memory effect}. 

Overall, Fig.~\ref{Fig8} provides a conceptual framework indicating which Reynolds-number and wall-distance regimes correspond to particular drag or wall corrections, and under which conditions additional effects such as lift, instationarities, or memory effects may occur.

The experimental results demonstrate the complexity of particle–wall interactions in low-Reynolds-number flows and highlight the limitations of existing stationary models. In particular, instationary and nonlinear effects observed at small Reynolds numbers in combination with elasticity cannot be captured by classic steady-state descriptions. The commonly used assumption that drag on a particle at is adequately described by stationary Stokes drag loses validity when walls are present within the Stokes length scale.

Finally, the experiments illustrate that even small variations in particle density, corresponding to Reynolds-number changes of order $10^{-1}$, can lead to qualitatively different dynamics in the creeping-flow regime. This regime is highly relevant to applications such as wastewater treatment involving microplastics, microorganism locomotion, blood flow, and particle transport in bioreactors. In such systems, sedimentation may dominate particle transport in regions of weak advection, making a detailed understanding of low-Reynolds-number sedimentation essential.

The development of improved models for particle motion in viscous fluids will therefore require the inclusion of fluid inertia (full Navier-Stokes equations), unsteady history effects and elastohydrodynamic interactions, despite their computational cost. A combination of targeted experiments, direct numerical simulations and data-driven approaches may ultimately enable the development of effective predictive models. As already noted by Stokes in 1851, even his classic formulation captures only part of the total resistance acting on a sphere, foreshadowing the need for more comprehensive descriptions of particle motion: "The formula ($-\textbf{\textit{F}}=6\pi\eta R \textbf{\textit{U}}$) determines, in the particular case of a sphere, that part of the whole resistance which depends on the first power of the velocity, even though the part which depends on the square of the velocity be not wholly insensible."\cite{Stokes.1851}

\backmatter

\bmhead{Supplementary information}
Available online along with the main manuscript.

\bmhead{Acknowledgements}
Funded by the Deutsche Forschungsgemeinschaft (DFG, German Research Foundation) – Project-ID 172116086 – SFB 926. The experiments and simulations were performed in frame of subproject A10 “Elastohydrodynamic interactions of particles flowing over microstructured surfaces” in the Collaborative Research Center 926 (CRC926 MICOS).

\section*{Declarations}

\begin{itemize}
\item Funding
DFG Project-ID 172116086 – SFB 926
\item Conflict of interest/Competing interests (check journal-specific guidelines for which heading to use)
The authors report no conclict of interest.
\item Ethics approval and consent to participate
Not applicable.
\item Data availability 
The datasets generated and analysed during the current study are available from the corresponding author on reasonable request..
\item Materials availability
Not applicable.
\item Code availability 
code used for data evaluation is available from the authors on reasonable request.

\end{itemize}

\noindent

\bigskip
\begin{flushleft}%
Editorial Policies for:

\bigskip\noindent
Springer journals and proceedings: \url{https://www.springer.com/gp/editorial-policies}

\bigskip\noindent
Nature Portfolio journals: \url{https://www.nature.com/nature-research/editorial-policies}

\bigskip\noindent
\textit{Scientific Reports}: \url{https://www.nature.com/srep/journal-policies/editorial-policies}

\bigskip\noindent
BMC journals: \url{https://www.biomedcentral.com/getpublished/editorial-policies}
\end{flushleft}

\begin{appendices}

\section{Additional analysis}\label{secA1}
\subsection{Additional analysis for Sec. \ref{sec4_1}}\label{secA1_1}
\paragraph{I) Detailed curve analysis for rigid spheres sedimenting at $\textbf{\textit{Re}}_\textbf{P}\mathbf{~\approx~}   \textbf{\textit{O}}\mathbf{(10^{-1})}$}
\textbf{I-a) Rigid sphere near rigid wall in the intermediate region} ($d/R\approx4$)\newline
Although there was a noticeable migration away from the wall of the rigid sphere which was released from $d/R\approx4$, there was no noticeable increase in velocity, \textit{cf.} purple curves in Fig.~\ref{Fig2}. The rigid sphere increased its wall distance by more than 20 \% of its radius ($\Delta d\geq0.2R$) over a sedimentation distance of $\approx40R$. A theoretical estimate of the increase in velocity resulting from the wall-induced lift of the nearest wall can be obtained by calculating the theoretical dimensionless velocities at the beginning and at the end of the experiment. As mentioned before, the influence of inertia on the drag of a sedimenting rigid sphere must be considered since the Reynolds number $\overline{Re}_\text{Peak}$ is of $O(10^{-1})$. For such a purpose, Faxén´s Eq. \ref{eq5} has been extended with additional Oseen terms of $O(Re)$.\cite{Tashibana.1976} The corrected dimensionless velocity for a sphere with larger mass inertia near a plane wall given by
\begin{equation}
    \label{eqA1}
{\left(\frac{U_\text{P}}{U_\text{th,ub}}\right)}_{\text{FO}, \left.\bullet \right|}=1-\frac{3}{16}Re_\text{P,th}-\frac{9}{16}\left(\frac{R}{d}\right)\cdot \chi+\frac{1}{8}\left(\frac{R}{d}\right)^3-\frac{45}{256}\left(\frac{R}{d}\right)^4-\frac{1}{16}\left(\frac{R}{d}\right)^5
\end{equation}
with
\begin{equation}
    \label{eqA1_2}
\chi=1-\frac{4}{3}\left(\frac{Re_\text{P,th}}{4\left(R/d\right)}\right)+\frac{23}{16}\left(\frac{Re_\text{P,th}}{4\left(R/d\right)}\right)^3-. . . .
\end{equation}
According to Eq. \ref{eqA1}, the rigid sphere with the corresponding material properties at $d/R\approx4$ would have a theoretical dimensionless velocity of approximately $\left.\left(U_\text{P}/U_\text{th,ub}\right)_\text{FO, $\bullet|$}\right|_{d/R=4}=0.86$. This theoretical value is, of course, larger than the measured value of about 0.8 since Eq. \ref{eqA1} considers only the influence of one wall on the velocity. Consequently, the velocity is overestimated. Nevertheless, Eq. \ref{eqA1} can be used to estimate a theoretical increase in velocity due to the wall-induced lift stemming from the closest wall.
The theoretical increase is calculated with the measured wall distances between positions $x/R\approx5$ (starting the migration) and $x/R\approx45$ (end of measurement). This results in a small linear increase of $\Delta_{5\ldots45}\left(U_\text{P}/U_\text{th,ub}\right)_\text{FO, $\bullet|$}\approx0.008$. The measured dimensionless velocity only shows an increase of approximately $\Delta_{5\ldots45}\left(U_\text{P}/U_\text{th,ub}\right)\approx0.002$ over the whole measurement range. Therefore, the dimensionless velocity appeared almost stationary after the transient acceleration in the experiments.\newline
\mbox{}\\
\textbf{I-b) Rigid sphere in the near-wall region }($d/R\approx2$)\newline
Interestingly, the increase in wall distance $\Delta d\approx0.1R$ at $d/R\approx2$ is smaller than in the intermediate region. However, the measured increase in velocity is larger than in the intermediate region ($\Delta_{5\ldots45}\left(U_\text{P}/U_\text{th,ub}\right)_\text{FO, $\bullet|$}\approx0.02$). This larger influence on the velocity is consistent with the theoretical estimations. Eq. \ref{eqA1} predicts a theoretical increase of $\Delta_{5\ldots45}\left(U_\text{P}/U_\text{th,ub}\right)_\text{FO, $\bullet|$}\approx0.012$. The theoretical dimensionless velocity of a sphere sedimenting near a plane wall at $d/R\approx2$ is $\left.\left(U_\text{P}/U_\text{th,ub}\right)_\text{FO, $\bullet|$}\right|_{d/R=2}=0.72$. The measured mean dimensionless velocity at $x/R\approx27$, \textit{i.e.}, where the sphere in the experiments again reached a distance of $d/R\approx2$ due to migration, is $U_\text{P,$\blacksquare$,2}/U_\text{th,ub}\approx0.7$.

\paragraph{II) Detailed curve analysis for soft spheres sedimenting at $\textbf{\textit{Re}}_\textbf{P}\mathbf{~\approx~}   \textbf{\textit{O}}\mathbf{(10^{-2})}$}
\textbf{II-a) Soft spheres in the intermediate region} ($d/R\approx4$)\newline
\indent \underline{\textit{Velocity:}} The soft spheres with the largest and the intermediate Young´s modulus ($E\approx1712$~kPa ($\blacktriangle$, blue) and $E\approx936$~kPa ($\blacklozenge$, green)) which were released from $d/R\approx4$ (Fig. \ref{Fig2}) accelerated due to their mass to a mean dimensionless velocity of approximately $U_\text{P}/U_\text{th,ub}\approx0.72\ldots0.73$. This was less than for the rigid sphere at this distance. After the initial acceleration due to its mass, the spheres sedimented with approximately constant velocity up to  $x/R\approx10$. After sedimentation of this distance, the spheres continued to accelerate, \textit{i.e.}, acceleration became apparent again. The soft sphere with largest $E$ accelerated up to a dimensionless velocity of $\left(U_\text{P,$\blacktriangle$}/U_\text{th,ub}\right)_\text{Max}\approx0.736$. This corresponds to an increase in velocity of $\Delta\left(U_\text{P,$\blacktriangle$}/U_\text{th,ub}\right)\approx0.015$ with respect to the first plateau. The spheres with intermediate $E$ accelerated up to $\left(U_\text{P,$\blacklozenge$}/U_\text{th,ub}\right)_\text{Max}\approx0.76$ ($\equiv \Delta \left(U_\text{P,$\blacklozenge$}/U_\text{th,ub}\right)\approx0.03$ with respect to the first plateau). Such an acceleration in multiple stages was also observed when the spheres sedimented in the duct center.\cite{Noichl.2022} Unlike with the release in the duct center, no second plateau with a terminal sedimentation velocity was established in the intermediate region.\newline
\indent The kinematics of sedimentation experiments of the softest spheres ($E\approx135$~kPa ($\CIRCLE$,orange)) which were released at $d/R\approx4$ showed a completely different, and, at first sight, surprising and unexpected behavior. After releasing the sphere, it accelerated up to a peak in dimensionless velocity of $\left(U_\text{P,$\CIRCLE$}/U_\text{th,ub}\right)_\text{Peak}\approx0.698$. 
Then, the sphere changed abruptly from acceleration to deceleration. The sphere decelerated until a minimum velocity of $\left(U_\text{P,$\CIRCLE$}/U_\text{th,ub}\right)_\text{Min}\approx0.66$ was reached. Interestingly, the dynamical behavior again changed at the sedimentation distance of $x/R\approx10$. From this dynamic switching point, the sphere accelerated again strongly up to $\left(U_\text{P,$\CIRCLE$}/U_\text{th,ub}\right)_\text{Max}\approx0.74$. The velocity curve is overall nonlinear. While such an unsteady and nonlinear behavior may seem surprising at first sight, it should be noted that the fact that sedimenting particles may accelerate and decelerate has already been observed experimentally in viscoelastic fluids, \textit{i.e.}, where further influence factors play a role.\cite{Becker.1994}

\indent \underline{\textit{Wall distance:}} Nonlinearities are also shown in the trajectories of the soft spheres (Fig. \ref{Fig2}, right side). In general, the trajectory is the curvier, the more deformable the spheres were. The inertial attraction phase of all soft spheres was followed by a repulsion phase, or migration, respectively. The smaller the Youngs´s modulus was, the steeper is the trajectory, \textit{i.e.}, the larger was the migration rate $d'(x)$. The trajectory of the soft sphere with largest $E$ (blue) flattens in the end. The intermediate soft and the softest sphere (green and orange) instead migrated to a maximum in wall distance and then turned back to the wall, \textit{i.e.}, they decreased the wall distance again.

If the trajectory of the softest sphere (orange) is compared with its velocity curve, the acceleration behavior becomes even more unintuitive. After the inertial wall attraction, the sphere decelerated although the wall distance increased strongly. Trajectory and velocity contradict each other from the perspective of wall-bounded models for the drag force like Faxén´s drag force model, see Eq. \ref{eq3}. In principle, the change from deceleration to acceleration would also suggest a change in direction of the trajectory. However, the trajectory shows an inflection point. At this point, there is a change from progressive increase to degressive increase in wall distance. This in turn suggests the decay of some kind of unsteady force at that point. Inflection points in the kinematics are already known from velocity curves during the first transient acceleration. A. B. Basset developed a decay function for the complete fall of a sphere starting from rest in an unbounded fluid. The resulting velocity curve is “S”-shaped and shows an inflection point when inertial forces such as inertia due to mass ($\textbf{\textit{F}}_\text{I}^\text{P}=m_\text{P} \frac{\text{d}\textbf{\textit{U}}_\text{P}}{\text{d}t}$) decay and drag becomes dominant.\cite{Basset.1888}, \cite{Tashibana.1976} The velocity curve of softest spheres which started sedimentation from $d/R\approx4$ shows an additional inflection point. However, the additional inflection point is not within the first transient acceleration, but at a much later time at position $x/R\approx26$. At this position, the trajectory shows a maximum in wall distance. From this position on, the sphere started to return to the wall, \textit{i.e.}, there is a second wall attraction phase. Meanwhile, the increase in velocity became degressive. It is already obvious from Fig. \ref{Fig2} that the shape of the velocity curves and trajectories continuously changes the softer the spheres become.

\textbf{II-b) Soft spheres in the near-wall region }($d/R\approx2$)\newline
\indent \underline{\textit{Velocity:}} Also, for the soft spheres released at a wall distance of $d/R\approx2$ , there is a clear influence that continuously emerges with increasing deformability or decreasing particle Reynolds number $Re_\text{P}$, respectively, see Fig. \ref{Fig3} (left side). As for the spheres released at a larger distance, there is an acceleration phase during which inertial wall attraction occurs, in the very beginning. The behavior that spheres decelerate after this first acceleration, which was already observed for the softest sphere released at the larger distance, is more prominent at shorter wall distance. Here, all soft spheres accelerated to a peak in velocity and then decelerated. The more deformable the spheres were and the lower the particle Reynolds number $Re_\text{P}$ was, the lower was the peak velocity. All peak velocities were lower than the velocity of the rigid spheres at that initial distance and the theoretical velocity calculated with Eq. \ref{eq3} $\left.\left(U_\text{P}/U_\text{St}\right)_\text{F, $\bullet|$}\right|_{d/R=2}=0.721$. Furthermore, the soft spheres had longer deceleration phases the more deformable they were and the lower the particle Reynolds number $Re_\text{P}$ was.\medskip \newline
\indent \underline{\textit{Wall distance:}} The softer the spheres were, the more nonlinear the migration became. The trajectory of the soft sphere with intermediate Young´s modulus flattens in the end. The trajectory of the softest sphere is qualitatively comparable with the one sedimenting in the intermediate region. The softest sphere decreased the wall distance again at the end although the dimensionless velocity remained almost constant in this phase.\medskip \newline
The soft sphere with largest $E$ ($\blacktriangle$, blue) decelerated slightly after the first acceleration phase (see Fig. \ref{Fig3}, left). Then, the sphere accelerated again to approximately the peak velocity. As with the rigid sphere, this could be related to the fact that drag is notably decreased by the sphere moving almost constantly away from the wall after the inertial wall attraction (see Fig. \ref{Fig3}, right). The migration rate $d'(x)$ was higher than the migration rate of the rigid spheres at that distance. The soft sphere with intermediate Young´s modulus ($\blacklozenge$, green) decreased the dimensionless velocity more than the stiffer one ($\Delta\left(U_\text{P,$\blacklozenge$}/U_\text{th,ub}\right)\approx-0.04$).
At position $x/R\approx20$ the sphere began to accelerate again. This sphere did not reach the peak velocity within the measurement range again. The velocity curves of spheres with the largest and intermediate Young´s modulus could possibly be described by catenaries, \textit{i.e.,} hyperbolic functions. The migration phase of spheres with intermediate $E$ started with a progressive increase in wall distance. Like in the experiments with this Young´s modulus starting from a larger initial distance, the trajectory shows an inflection point at $x/R\approx10$. There, the migration went over into degressive increase.\newline 
\indent The softest sphere ($\CIRCLE$, orange) which started from $d/R\approx2$ decreased its dimensionless velocity most. The peak in dimensionless velocity after the mass acceleration phase $\left(U_{\text{P},\CIRCLE}/U_\text{th,ub} \right)_\text{Peak}\approx0.6$ was reduced to $\left(U_{\text{P},\CIRCLE}/U_\text{th,ub}\right)_\text{Min}\approx0.52$, \textit{i.e.,} a total decrease of $\Delta\left(U_{\text{P},\CIRCLE}/U_\text{th,ub}\right)\approx-0.08$. 

\begin{figure}
    \centering
    \includegraphics[width=\textwidth]{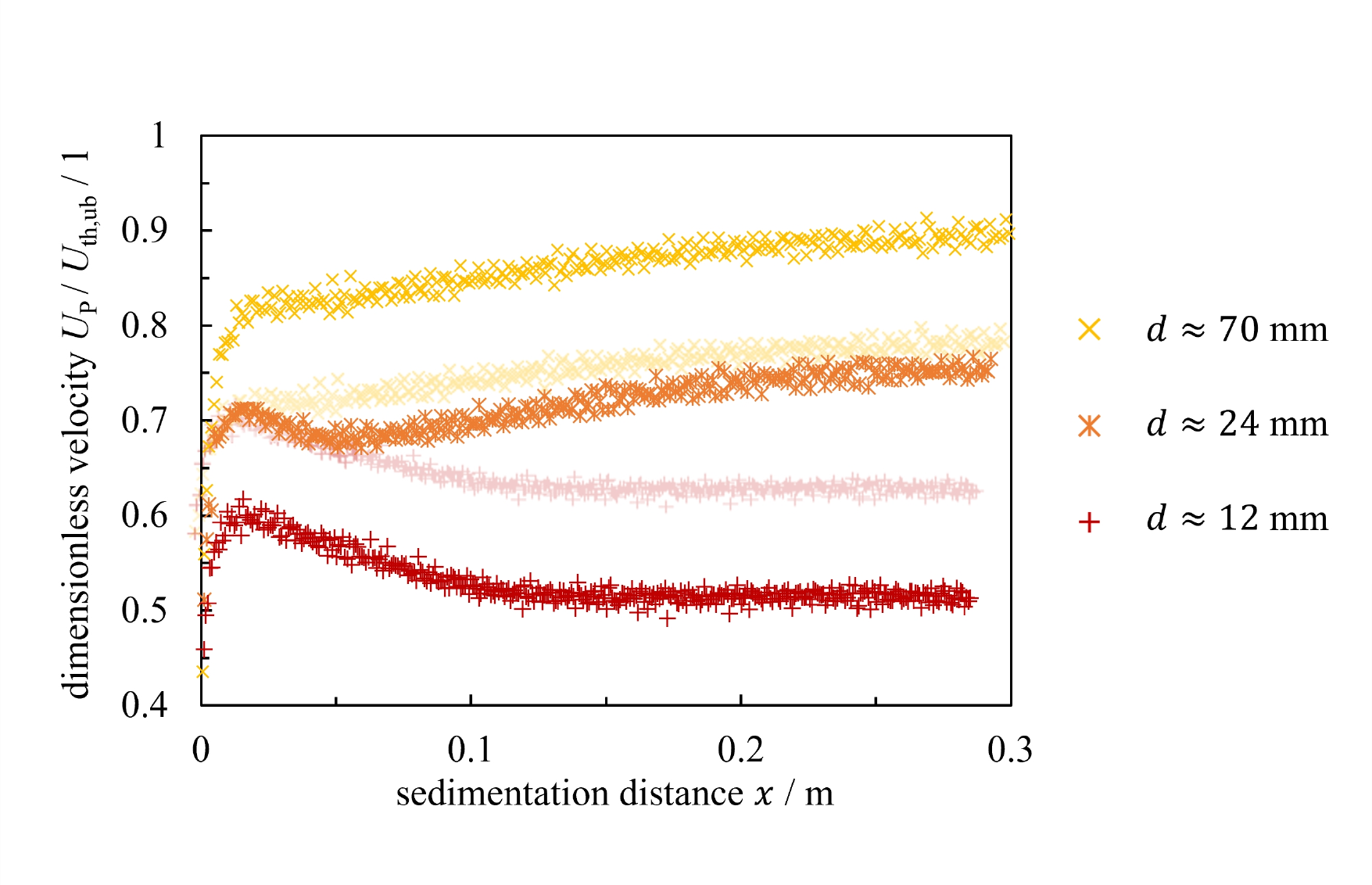}
    \caption{Dimensionless particle velocity $U_\text{P}/U_\text{th,ub}$ plotted over the sedimentation distance $x$ of sedimenting soft spheres ($E\approx135$~kPa) with a radius of $R\approx6$~mm in the vicinity of a plane, rigid wall. The spheres started from various wall distances of $d\approx12$~mm ($d/R\approx2$); $d\approx24$~mm ($d/R\approx4$) and $d\approx70$~mm (center of a rectangular duct); Bright curves: upwards and downwards shifted curves, matched on the point of the peak velocity of the intermediate curve.}
    \label{FigA1}
\end{figure}

\subsection{Additional analysis for Sec. \ref{sec4_2} - Detailed curve analysis for rigid spheres sedimenting at $Re_\text{P}\approx O(10^{-2})$}\label{secA1_2}

\indent \underline{\textit{Velocity:}} The small, rigid sphere accelerated to a peak in velocity of $\left(U_\text{P,$\blacksquare$}/U_\text{th,ub}\right)_\text{Peak} \approx 0.61$. The dimensionless peak velocity is in the same range as the $\left(U_\text{P}/U_\text{th,ub}\right)_\text{Peak}$ of the softest sphere which had a comparable Reynolds number. The fact that, at the same particle Reynolds number, both the small rigid sphere as well as the larger soft sphere accelerated to the same reduced peak velocity and then decelerated implies that the velocity reduction after the first transient phase is a hydrodynamics-dependent quantity and not an elasticity-induced effect. After reaching the peak, the small sphere decelerated in good approximation linearly to a dimensionless velocity of $\left(U_\text{P,$\blacksquare$}/U_\text{th,ub}\right)_\text{Min}\approx0.57$, \textit{i.e.,} a total decrease of $\Delta \left(U_\text{P,$\blacksquare$}/U_\text{th,ub}\right)\approx-0.04$. This decrease is approximately half the decrease of the softest sphere. This in turn implies that elasticity enhanced deceleration, \textit{i.e.,} elasticity led to a larger deceleration rate. Thereafter, the small sphere sedimented stationary with the reduced velocity $\left(U_\text{P,$\blacksquare$}/U_\text{th,ub}\right)_\text{Min}$.\medskip \newline
\indent \underline{\textit{Wall distance:}}
The initial dimensionless wall distance of the small sphere in the experiment was $d/R\approx2.25$. This is initially further away from the wall than the experiments used for the comparison (in relation to its radius---not in absolute terms). However, experimentally, it became more difficult to set the initial distance accurately the smaller the spheres were. Like all previously considered experiments, the small sphere reduced the wall distance after releasing it from the pipette. The inertial wall attraction was also observable during the first transient acceleration phase. However, the total decrease in wall distance $\Delta \left(d/R\right)_{R=4mm}=\left(d/R\right)_0-\left(d/R\right)_\text{min}\approx-0.01R$ due to the inertial wall attraction was much smaller than for the larger spheres ($\Delta\left(d/R\right)_{R=6mm}\approx-0.04R$). After the inertial wall attraction, the sphere increased the wall distance progressively and changed quickly to a degressive increase. After a sedimentation distance of approximately 40R was reached, the trajectory flattens. From there on, the sphere sedimented at an equilibrium wall distance without further migration. Mapping the trajectories on each other at the point of minimum wall distance shows that the total increase $\Delta(d/R)=\left(d/R\right)_\text{max}-\left(d/R\right)_\text{min}$ is of the same order of magnitude (see right diagram in Fig. \ref{Fig5}). Unlike the soft spheres, the small rigid spheres had no second attraction phase at the end. Additionally, the trajectory of the soft sphere shows larger curvature during the migration phase compared to the small, rigid sphere.\medskip \newline

\begin{figure}
    \centering
    \includegraphics[width=\textwidth]{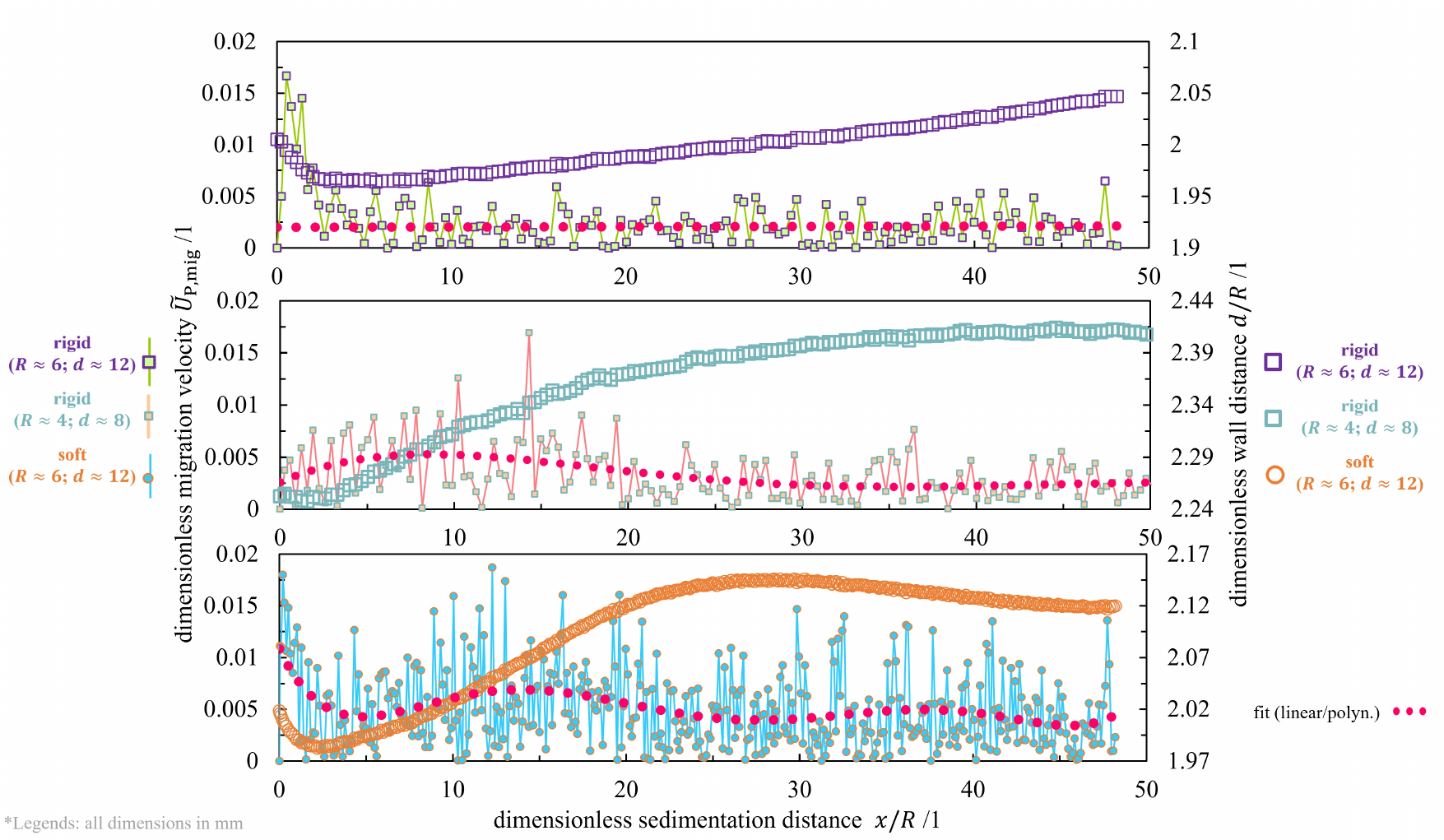}
    \caption{Dimensionless migration velocity $\widetilde{U}_\text{P,mig}$ of a rigid sphere ($R\approx4$~mm and $R\approx6$~mm) and a soft sphere ($R\approx6$~mm; $E=135$~kPa) sedimenting near a plane, rigid wall plotted over the dimensionless sedimentation distance $x/R$ (filled symbols with lines). Red, dotted curves: linear fit curve (top) and polynomial fits with $n=5$ (middle) and $n=6$ (bottom) of the measured migration velocity; Hollow symbols: corresponding trajectory ($\left(x/R\right) - \left(d/R\right)$-plot); The spheres started from an initial position of $d/R\approx2$.}
    \label{FigA2}
\end{figure}

\indent \underline{\textit{Migration velocity:}}
The curvature in the trajectory is directly related to the migration velocity $U_\text{P,y}$. $U_\text{P,y}$ is the velocity directed perpendicularly to the sedimentation velocity. Fig.~\ref{FigA2} shows measurements of the dimensionless migration velocity $\widetilde{U}_\text{P,mig}=U_\text{P,y}/U_\text{th,ub}$ of a large, rigid sphere ($R\approx6$~mm) with $Re_\text{P}\approx{10}^{-1}$(top), a small, rigid sphere ($R\approx4$~mm) with $Re_\text{P}\approx{10}^{-2}$(middle) and a soft sphere ($R\approx6$~mm) with $Re_\text{P}\approx{10}^{-2}$(bottom). All experiments were performed with an initial wall distance of $d/R\approx2$. The corresponding trajectories are also plotted in the diagram on the secondary axis.\newline

The migration velocity curve of the large, rigid sphere ($R\approx6$~mm; $Re_\text{P}\approx O(10^{-1})$; top diagram) is still in line with expectations. The migration velocity is small and scatters around a constant value. Consistently, linear migration leads to a stationary migration velocity ($\widetilde{U}_\text{P,mig}=U_\text{P,y}/U_\text{th,ub}\approx0.0021 \pm 0.0015$, \textit{i.e.}, 0.2~\% of the theoretical sedimentation velocity in an unbounded fluid $U_\text{th,ub}$, \textit{cf.} Tab.\ref{tab3}). During the initial acceleration phase (or the inertial wall attraction, respectively), there is a peak in velocity. The peak velocity is several times larger than the constant migration velocity. After the inertial attraction was finished, the velocity decreased abruptly. The velocity curve of the small, rigid sphere ($R\approx4$~mm; $Re_\text{P}\approx O(10^{-2})$; middle) shows more fluctuations of the velocity during the migration phase. The maximum of $\widetilde{U}_\text{P,mig}$ was during the migration phase and was approximately 0.5 \% of $U_\text{th,ub}$ (cf. the fit curve). The maximum is located approximately at the inflection point which appears at a sedimentation distance of $\approx10R$. After the maximum was reached, the velocity decreased to a small and constant velocity which is comparable with the dimensionless, terminal migration velocity of the large sphere. 

The migration velocity of the soft spheres ($R\approx6$~mm; $Re_\text{P}\approx O(10^{-2})$; bottom) fluctuated strongly over the entire measurement range. For the softest sphere, a peak in migration velocity during the inertial wall attraction phase was detected. The $\widetilde{U}_\text{P,mig}$ in this phase is of same order of magnitude as in the experiments with the large, rigid spheres. The velocities during the migration phase were overall larger than for the small, rigid spheres. The maximum in the mean migration velocity (\textit{cf.} the fit curve) is also located at the inflection point which appears at a sedimentation distance of $\approx10R$. This corresponds to the inflection point in the trajectory. Unlike the rigid spheres, no constant velocity after the migration phase was detected for the soft spheres. The dimensionless migration velocity after the migration phase was in general larger ($\widetilde{U}_\text{P,mig}=U_\text{P,y}/U_\text{th,ub}\approx0.0049 \pm 0.0038$ for $x/R>10$). Additionally, a second peak in velocity was detected during the second attraction phase. However, the second peak was of smaller order of magnitude as the velocities during the inertial attraction and during the migration phase (\textit{cf.} polynomial fit curve).\newline
\section{Additional information for Simulations from Sec.~\ref{sec4_3}}\label{secB}
\begin{table}
    \centering
    \renewcommand{\arraystretch}{2.5}
    \begin{tabular}{ c c c c c }
        {$d$ /mm} &  $\rho_{\text{s}}$ / kg$\cdot$m\textsuperscript{-3} & $U_\text{P, exp}$ /mm$\cdot$s\textsuperscript{-1} & $U_\text{P, sim}$ /mm$\cdot$s\textsuperscript{-1} & $\Delta_\text{sim}$\\
        \hline
        
        70 & 1159.21 & 11.9 & 12.60 & +5.9 \% \\
        70 & 987.59 & \makecell{1.1 (1st plateau) \\ 1.2 (2nd plateau)} & 1.10 & -0.4 \% \\
         12 & \makecell{ 1159.8 (Sim) \\ 1159.85 (Exp)} & 9.9  & 10.73 & +8.1 \% \\
         12 & \makecell{ 987.59 (Sim) \\ 988.24 (Exp)} & 0.8 (Peak)  & 0.93 & +16.5 \% \\
    \end{tabular}
     \caption{Simulated sedimentation velocity for solid densities $\rho_{\text{s}}\approx 988$~kg$\cdot$m\textsuperscript{-3}  and $\rho_{\text{s}}\approx 1159$~kg$\cdot$m\textsuperscript{-3} in the center and in the near-wall region of a rectangular duct }
    \label{tabB1}
\end{table}

\textbf{Experimental vs. simulated dimensionless sedimentation velocities in the near-wall region}\medskip \newline
$\left(U_\text{P,exp}/U_\text{th,ub}\right)_{d=12; 1159.85}\approx 9.9/14.2\approx 0.70$\newline
$\left(U_\text{P,sim}/U_\text{th,ub}\right)_{d=12; 1159.8}\approx 10.7/14.6\approx 0.73$\newline
$\left(U_\text{P,exp,Peak}/U_\text{St}\right)_{d=12; 988.24}\approx 0.8/1.3\approx 0.61$\newline
$\left(U_\text{P,sim}/U_\text{St}\right)_{d=12; 987.59}\approx 0.93/1.31\approx 0.71$\newline

\begin{figure}
    \centering
    \includegraphics[width=0.5\textwidth]{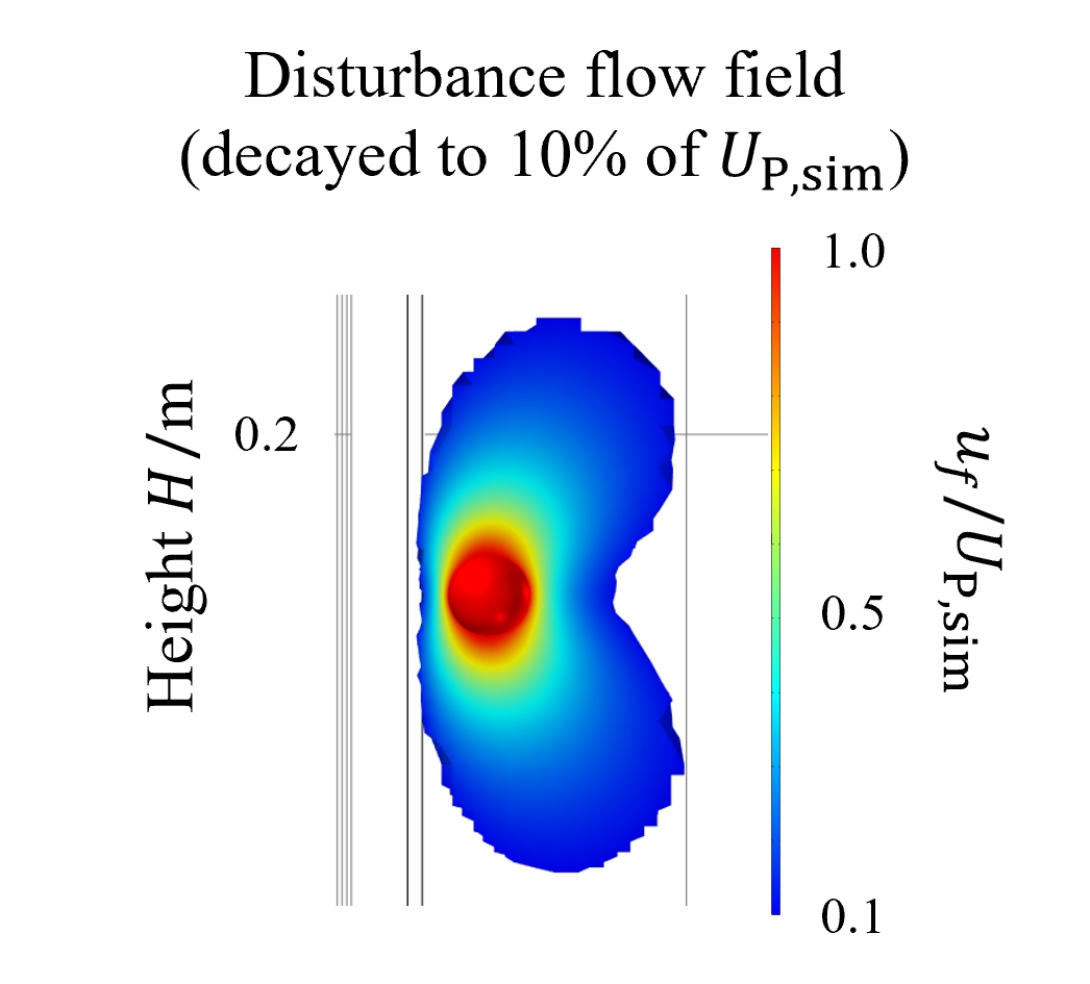}
    \caption{Simulated disturbance flow field decaying to 10\% of $U_\text{P,sim}$ of a rigid sphere ($R=6$~mm, $\rho_{\text{s}}=987.59$~kg$\cdot$m\textsuperscript{-3}) sedimenting at $Re_\text{P}\approx O(10^{-2})$ at a distance $d=12$~mm from a no-slip boundary in a rectangular fluid domain.}
    \label{FigA2_Disturbance}
\end{figure}




\end{appendices}


\bibliography{sn-bibliography}

\end{document}